\documentclass[conference]{IEEEtran}
\IEEEoverridecommandlockouts
\usepackage{cite}
\usepackage{color}
\usepackage{amsmath,amssymb,amsfonts}
\usepackage{algorithmic}
\usepackage{graphicx}
\usepackage{textcomp}
\def\BibTeX{{\rm B\kern-.05em{\sc i\kern-.025em b}\kern-.08em
    T\kern-.1667em\lower.7ex\hbox{E}\kern-.125emX}}

\usepackage[ruled]{algorithm2e}

\usepackage{url}

\def\BibTeX{{\rm B\kern-.05em{\sc i\kern-.025em b}\kern-.08em T\kern-.1667em\lower.7ex\hbox{E}\kern-.125emX}}

\newcommand {\mymarginpar}[1]{\marginpar{#1}}
\renewcommand {\marginpar}[1]{}

\def\_{\rule{.3em}{.15ex}}      

\newcommand{\ls}[1]
   {\dimen0=\fontdimen6\the\font
    \lineskip=#1\dimen0
    \advance\lineskip.5\fontdimen5\the\font
    \advance\lineskip-\dimen0
    \lineskiplimit=.9\lineskip
    \baselineskip=\lineskip
    \advance\baselineskip\dimen0
    \normallineskip\lineskip
    \normallineskiplimit\lineskiplimit
    \normalbaselineskip\baselineskip
    \ignorespaces
   }

\newcommand {\bearn}{\begin{eqnarray*}}
\newcommand {\eearn}{\end{eqnarray*}}
\newcommand {\barr}{\begin{array}}
\newcommand {\earr}{\end{array}}

\newcommand {\N}{{\cal N}}

\newtheorem{definition}{Definition}
\newtheorem{property}[definition]{Property}
\newtheorem{proposition}[definition]{Proposition}
\newtheorem{lemma}[definition]{Lemma}
\newtheorem{theorem}[definition]{Theorem}
\newtheorem{corollary}[definition]{Corollary}
\newtheorem{example}{Example}
\newtheorem{remark}[definition]{Remark}

\newcommand {\benum} {\begin{enumerate}}
\newcommand {\eenum} {\end{enumerate}}

\newcommand {\bdesc} {\begin{description}}
\newcommand {\edesc} {\end{description}}

\newcommand {\bfig}[2] {\begin{figure}
  \centering
  \includegraphics[width=#2]{#1}}
\newcommand {\brotatefig}[2] {\begin{figure}[htbp]
                        \centerline {
                         \epsfig{figure={#1},clip=,angle=-90,width={#2}}}}
\newcommand {\bfigfirst}[2] {\begin{figure}[h]
                        \centerline {
                        \setlength{\epsfxsize}{#2}
                        \epsffile{#1}}}
\newcommand {\efig}[2]{ \caption{#2}
                        \label{fig:#1}
                        \end{figure}
                        \mymarginpar{fig:#1}}
\newcommand {\erotatefig}[2]{ \caption{#2}
                        \label{fig:#1}
                        \end{figure}
                        \mymarginpar{fig:#1}}
\newcommand {\rfig}[1]{Figure \ref{fig:#1}}

\newcommand {\btab}[1]{
                       \begin{table}
                       \centering
                       \begin{tabular}{#1}}
\newcommand {\etab}[3] {
                       \end{tabular}
                       \caption[#3]{#2}
                       \label{tab:#1}
                       \end{table}
                       \mymarginpar{tab:#1}
                       \vspace{.1in}}

\newcommand {\btabular}[1]{\begin{center}
                       \begin{tabular}{#1}}
\newcommand {\etabular}{\end{tabular}
                       \end{center}}

\newcommand {\bdefin}[1]{\begin{definition}
                      \mymarginpar{def:#1}
                      \label{def:#1} }
\newcommand {\edefin}       {\end{definition}}
\newcommand {\rdef}[1]{Definition \ref{def:#1}}

\newcommand {\bpro}[1]{\begin{property}
                      \mymarginpar{pro:#1}
                      \label{pro:#1} }
\newcommand {\epro}   {\end{property}}

\newcommand {\bprop}[1]{\begin{proposition}
                      \mymarginpar{prop:#1}
                      \label{prop:#1} }
\newcommand {\eprop}       {\end{proposition}}
\newcommand {\rprop}[1]{Proposition \ref{prop:#1}}

\newcommand {\blem}[1]{\begin{lemma}
                      \mymarginpar{lem:#1}
                      \label{lem:#1} }
\newcommand {\elem}   {\end{lemma}}
\newcommand {\rlem}[1]{Lemma \ref{lem:#1}}

\newcommand {\bthe}[1]{\begin{theorem}
                      \mymarginpar{the:#1}
                      \label{the:#1} }
\newcommand {\ethe}   {\end{theorem}}
\newcommand {\rthe}[1]{Theorem \ref{the:#1}}

\newcommand {\bproof}{\noindent {\bf Proof.} \ }
\newcommand {\eproof} {\hfill \squares \\ \vspace{.3cm}}
\newcommand {\bcor}[1]{\begin{corollary}
                      \mymarginpar{cor:#1}
                      \label{cor:#1} }
\newcommand {\ecor}   {\end{corollary}}

\newcommand {\bax}[1]{\begin{axiom}
                      \mymarginpar{ax:#1}
                      \label{ax:#1} }
\newcommand {\eax}       {\vspace{-.1in} \end{axiom}}

\newcommand {\bex}[2]{\vspace{.1in}
                      \begin{example}
                      \mymarginpar{ex:#1}
                       {\bf #2}
                      \label{ex:#1} }
\newcommand {\eex}       {\end{example} \vspace{.3cm} }
\newcommand {\rex}[1]{Example \ref{ex:#1}}

\newcommand {\brem}[1]{\begin{remark}
                      \mymarginpar{rem:#1}
                      \label{rem:#1} \em }
\newcommand {\erem}   {\end{remark}}

\newcommand {\beq}[1]{\mymarginpar{eq:#1}
                      \begin{equation}
                      \label{eq:#1} }

\newcommand {\beqno}[1]{\mymarginpar{eq:#1}
                      \begin{eqnarray}
                      \nonumber}

\newcommand {\eeq}       {\end{equation}}
\newcommand {\eeqno}       { && \end{eqnarray}}
\newcommand {\req}[1]{(\ref{eq:#1})}

\newcommand {\bear}[1]{\mymarginpar{eq:#1}
                       \begin{eqnarray}
                       \label{eq:#1} }

\newcommand {\bearno}[1]{\mymarginpar{eq:#1}
                       \begin{eqnarray}
                       \nonumber}

\newcommand {\eear}{\end{eqnarray}}
\newcommand {\eearno}{\end{eqnarray}}
\newcommand {\bsel}{\left \{ \begin{array}{cl}}
\newcommand {\esel}{\end{array} \right.}

\newcommand {\bmat}[1]{\left [ \begin{array}{#1}}
\newcommand {\emat}{\end{array} \right ]}
\newcommand {\bsec}[2]{\mymarginpar{sec:#2}
                       \section{#1}
                       \label{sec:#2} }

\newcommand {\rsec}[1]{Section \ref{sec:#1}}

\newcommand {\bsubsec}[2]{\mymarginpar{sec:#2}
                       \subsection{#1}
                       \label{sec:#2} }

\def\R{I\kern-0.30em R}
\def\N{I\kern-0.30em N}
\def\P{I\kern-0.30em P}
\newcommand\squares{\vrule height6pt width7pt depth1pt}

\def\pr{{\bf\sf P}}

\newcommand{\rhog}{\rho}
\newcommand{\rone}{\tau}
\newcommand{\finc}{{\cal F}^\uparrow}

\begin{document}

\title{ALOHA Receivers: a Network Calculus Approach for Analyzing Coded Multiple Access with SIC}

\author{Tzu-Hsuan Liu, Che-Hao Yu, Yi-Jheng Lin, Cheng-Shang~Chang,~\IEEEmembership{Fellow,~IEEE,}\\ and Duan-Shin Lee,~\IEEEmembership{Senior Member,~IEEE}
                \thanks{T.-H. Liu, C.-H. Yu, Y.-J. Lin, C.-S. Chang, and D.-S. Lee are with the Institute of Communications Engineering, National Tsing Hua University, Hsinchu 30013, Taiwan, R.O.C. Email: carina000314@gmail.com; chehaoyu@gapp.nthu.edu.tw; s107064901@m107.nthu.edu.tw;   cschang@ee.nthu.edu.tw;  lds@cs.nthu.edu.tw.  This work was supported in part by the Ministry of
Science and Technology, Taiwan, under Grant 109-2221-E-007-091-MY2, and in part by Qualcomm Technologies under Grant SOW NAT-435533.}
}

\maketitle
\begin{abstract}
Motivated by the need  to hide the complexity of the physical layer from performance analysis in a layer 2 protocol, a class of abstract receivers, called Poisson receivers, was recently proposed in \cite{chang2020Poisson} as a probabilistic framework for providing differentiated services in  uplink transmissions in 5G networks.
In this paper, we further propose a deterministic framework of ALOHA receivers that can be incorporated into the probabilistic framework of Poisson receivers for
analyzing coded multiple access with successive interference cancellation. An ALOHA receiver is characterized by a success function of the number of packets that can be successfully received. Inspired by the theory of network calculus, we derive various algebraic properties for several operations on success functions and use them to prove
various closure properties of ALOHA receivers, including (i) ALOHA receivers in tandem,
(ii) cooperative ALOHA receivers, (iii) ALOHA receivers with traffic multiplexing, and (iv) ALOHA receivers with packet coding.
By conducting extensive simulations, we show that our theoretical results  match extremely well with the simulation results.

\end{abstract}





%

\bsec{Introduction}{introduction}

The fifth-generation networks (5G) and beyond aim to cover
three generic connectivity types: (i) enhanced mobile broadband (eMBB), (ii) ultra-reliable low-latency communications (URLLC), and (iii) massive machine-type communications (mMTC) (see, e.g., \cite{li20175g,bennis2018ultra,popovski2019wireless} and references therein). The reliability defined in 3GPP for supporting URLLC services, such as autonomous driving, drones, and augmented/virtual reality, requires the $1-10^{-5}$ success probability of transmitting a layer 2 protocol data unit of 32 bytes within 1ms. On the other hand,
 the number of devices is in general much larger than the number of orthogonal resources for mMTC. Motivated by these emerging needs in 5G, many multiple access schemes have been proposed in the literature, see, e.g., Contention Resolution Diversity Slotted ALOHA (CRDSA) \cite{casini2007contention}, Irregular Repetition Slotted
ALOHA (IRSA) \cite{liva2011graph}, coded slotted ALOHA (CSA) \cite{narayanan2012iterative,paolini2012random,jakovetic2015cooperative,stefanovic2018coded},
 Low-Density Signature (LDS) based spreading \cite{Hoshyar2008}, Sparse Code Multiple Access (SCMA) \cite{SCMA},
 Multi-User Sharing Access (MUSA) \cite{Yuan2016}, Pattern Division Multiple Access (PDMA) \cite{Chen2017},
 $T$-fold ALOHA \cite{ordentlich2017low}, Asynchronous Multichannel Transmission Schedules (AMTS) \cite{chang2019asynchronous}, Grant-free Hybrid Automatic Repeat reQuest
(GF-HARQ) \cite{liu2020analyzing}, and
 Polar Code Based TIN-SIC \cite{andreev2020polar}. Though many of the above multiple access schemes
 use elegant message passing algorithms (or density evolution methods) for decoding as in the low-density parity-check codes \cite{gallager1962low}, the decoding results
 depend heavily on the performance of the underlining physical channel. As such, it is very difficult to carry out further analysis in a layer 2 protocol, where there are multiple classes of input traffic and interconnected base stations/receivers/relays/antennas.

 To hide the complexity from the physical layer, one needs an abstract model at the Medium Access Control (MAC) layer. Motivated by this, we proposed in our previous work \cite{chang2020Poisson} an abstract receiver, called {\em Poisson receiver}, that models the input-output relations of multiple classes of input traffic subject to an (independent) Poisson offered load. It was shown in \cite{chang2020Poisson}
that  various CSA systems in \cite{liva2011graph,narayanan2012iterative,paolini2012random,jakovetic2015cooperative,stefanovic2018coded} can be modelled by Poisson receivers. In this paper, we take one step further to show that  the block fading channel with capture in \cite{stefanovic2014exploiting,clazzer2017irregular,stefanovic2018coded} can also be modelled by Poisson receivers.
 Poisson receivers have two elegant closure properties: (i) Poisson receivers with {\em packet routing} are still Poisson receivers, and
(ii) Poisson receivers with  {\em packet coding} are still Poisson receivers.
Thus, one can use smaller Poisson receivers as building blocks for analyzing a larger Poisson receiver.
Intuitively, Poisson receivers can be viewed as a probabilistic framework for analyzing coded random access.

In this paper, we propose a
new class of abstract receivers, called {\em ALOHA receivers},
that can be viewed as a deterministic framework for analyzing coded multiple access with successive interference cancellation (SIC).
Our approach is inspired by the theory of network calculus (see, e.g., the seminal works by Rene Cruz \cite{cruz1991calculus,cruz1991calculusb}, the books and the survey papers \cite{chang2012performance,le2001network,mao2006survey,jiang2008stochastic,fidler2010survey} and references therein).
The theory of network calculus is a queueing theory that analyzes the network performance by viewing each queue as a building block with a certain arrival-departure property. By exploiting various elegant algebraic properties of the min-plus algebra \cite{baccelli1992synchronization}, end-to-end delay bounds can be derived for various scheduling policies, including Generalized Processor Sharing (GPS) \cite{parekh1993generalized} and Service Curve Earliest Deadline first scheduling (SCED) \cite{sariowan1999sced}.

Like a Poisson receiver, an ALOHA receiver is also an abstract receiver that views the underlining physical layer as a network element with a {\em deterministic} input-output function. Specifically, when there are $n_k$ class $k$ packets, $k=1,2,\ldots,K$, arriving at a $\phi$-ALOHA receiver,  the number of class $k$ packets that are successfully received (or successfully decoded) is exactly $\phi_k(n)$, $k=1,2, \ldots, K$. The function $\phi(n)=(\phi_1(n), \phi_2(n), \ldots, \phi_K(n))$ is called the {\em success} function of the $\phi$-ALOHA receiver when it is subject to a {\em deterministic} load   $n=(n_1, n_2, \ldots, n_K)$. On the other hand, the {\em failure} function, denoted by $\phi^c$
with $\phi^c(n)= n-\phi(n)$ represents the number of packets remained to be decoded.

One nice feature of coded multiple access with SIC is that a user can send multiple copies of a packet to a receiver.
As long as one of the copies is successfully received, then the packet is successfully received, and the other copies of that packet can be removed from the receiver. Such a feature is known as the perfect SIC assumption in the literature (see, e.g., \cite{stefanovic2018coded,andreev2020polar}). As such, one can repeatedly apply the SIC operation to a receiver until there are no more packets that can be successfully decoded. Such an operation is even more interesting and powerful when a set of cooperative receivers can exchange information of successfully received packets.

The SIC operation induces several operations on functions, including minimum, composition, closure, and complement. Like the theory of network calculus, we develop various algebraic properties for these operations that can be used for proving various closure properties of ALOHA receivers, including
\begin{description}
\item[(i)] two ALOHA receivers in tandem is also an ALOHA receiver in \rsec{tandem},
\item[(ii)] two cooperative receivers is an ALOHA receiver in \rsec{coop},
\item[(iii)] ALOHA receivers with traffic multiplexing is an ALOHA receiver in \rsec{multiplex},
\item[(iv)] ALOHA receivers with packet coding is an ALOHA receiver in \rsec{codingALOHA}, and
\item[(v)] multiple cooperative $D$-fold ALOHA receivers is an ALOHA receiver in \rsec{SAmul}.
\end{description}

An ALOHA receiver can be easily converted into a Poisson receiver. However, it is not the other way around.
As such, the deterministic framework of ALOHA receivers can be incorporated into the probabilistic framework of Poisson receivers. To illustrate this, we provide a numerical example that uses two cooperative $D$-fold ALOHA receivers with packet coding to provide differentiated services between URLLC traffic and eMBB traffic. For $D=1$, such a system is reduced to the CSA system with two cooperative receivers in \cite{chang2020Poisson}.
The theoretical results in this example match extremely well with the simulation results (except for those data points with extremely small error probabilities). Moreover, there is a significant performance gain by using the $2$-fold ALOHA receivers over the $1$-fold ALOHA receivers.

\begin{figure}[ht]
	\centering
	\includegraphics[width=0.43\textwidth]{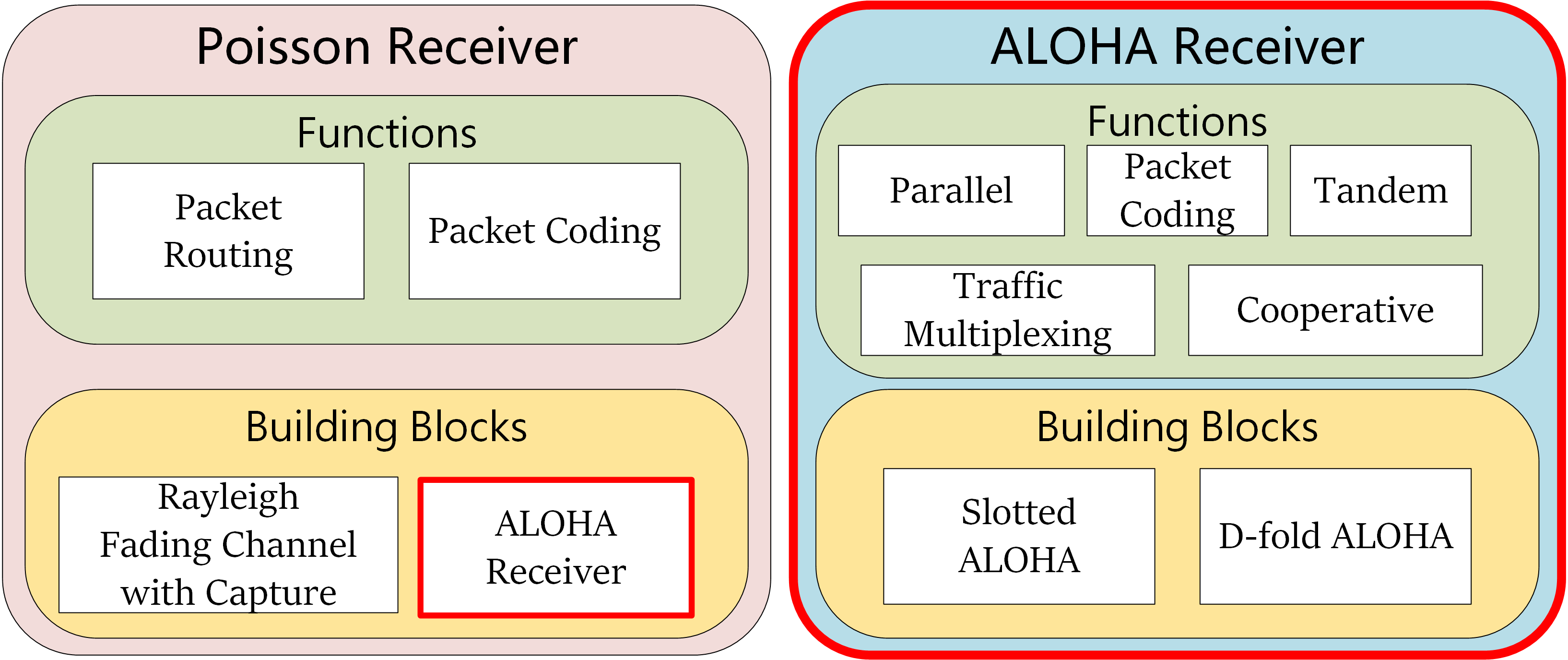}
	\caption{An overview of the probabilistic framework of Poisson receivers and the deterministic framework of ALOHA receivers.}
	\label{fig:overview}
\end{figure}

To provide a general overview of the probabilistic framework of Poisson receivers and the deterministic framework of ALOHA receivers, we depict in \rfig{overview} the basic building blocks and the associated calculus for analyzing coded multiple access with SIC.

The rest of the paper is organized as follows.
In \rsec{poisson}, we briefly review the framework of Poisson receivers in our previous work \cite{chang2020Poisson}.
There we also show how to model the Rayleigh  block fading channel with capture as a Poisson receiver.
We develop the deterministic framework of ALOHA receivers in \rsec{phiALOHA} by proving various algebraic properties for several operations on functions. Using these algebraic properties, we then prove various closure properties of ALOHA receivers. In \rsec{num}, we provide numerical examples to show how ALOHA receivers and Poisson receivers can be used for
providing differentiated services between URLLC traffic and eMBB traffic.
The paper is then concluded in \rsec{con}.

\bsec{Poisson receivers}{poisson}

\bsubsec{Review of the framework of Poisson receivers}{def}

In this section, we briefly review the framework of Poisson receivers in our previous work \cite{chang2020Poisson}.
We say a system with $K$ classes of input traffic is subject to a Poisson offered load  $\rho=(\rho_1, \rho_2, \ldots, \rho_K)$ if these $K$ classes of input traffic are {\em independent}, and the number of class $k$ packets arriving at the system
follows a Poisson distribution with mean $\rho_k$, for $k=1,2, \ldots, K$.


\bdefin{Poissonmul}{(\bf Poisson receiver with multiple classes of input traffic \cite{chang2020Poisson})}
An abstract receiver
is called a {\em $(P_{{\rm suc},1}(\rhog), P_{{\rm suc},2}(\rhog), \ldots, P_{{\rm suc},K}(\rhog))$-Poisson receiver} with $K$ classes of input traffic if the receiver is subject to a Poisson offered load  $\rho=(\rho_1, \rho_2, \ldots, \rho_K)$, a tagged (randomly selected) class $k$  packet
is successfully received with probability $P_{{\rm suc},k}(\rhog)$, for $k=1,2, \ldots, K$.
\edefin

The throughput of class $k$ packets  (defined  as the expected number of class $k$ packets that are successfully received) for a {\em $(P_{{\rm suc},1}(\rhog), P_{{\rm suc},2}(\rhog), \ldots, P_{{\rm suc},K}(\rhog))$-Poisson receiver} subject to a Poisson offered load $\rho$ is thus
\beq{Poithrmul}
S_k=\rho_k \cdot P_{{\rm suc},k}(\rhog),
\eeq
$k=1,2, \ldots, K$.

It was shown in \cite{chang2020Poisson} that many systems can be modelled by Poisson receivers, including SA, SA with multiple {\em non-cooperative} receivers, and SA with multiple {\em cooperative} receivers.
Moreover, there are two elegant closure properties of Poisson receivers in \cite{chang2020Poisson}.
\begin{description}
\item[(i)]  Poisson receivers with {\em packet routing} are still Poisson receivers.
\item[(ii)] Poisson receivers with  {\em packet coding} are still Poisson receivers.
\end{description}
These two closure properties allow us to use smaller Poisson receivers as building blocks for analyzing a larger Poisson receiver.

\bthe{routing}{\bf (Poisson receivers with packet routing \cite{chang2020Poisson})}
Consider a Poisson receiver with $K_2$ classes of input traffic and the success probability functions $P_{{\rm suc},1}(\rhog), P_{{\rm suc},2}(\rhog), \ldots, P_{{\rm suc},K_2}(\rhog)$.
There are $K_1$ classes of {\em external} input traffic to the Poisson receiver.
With the routing probability $r_{k_1, k_2}$,
a class $k_1$ external packet transmitted to the Poisson receiver becomes a class $k_2$ packet at the Poisson receiver.
Let $G=(G_1, G_2, \ldots, G_{K_1})$ and $\rho=(\rho_1,\rho_2, \ldots, \rho_{K_2})$ with
\beq{mean4444rou}
\rho_{k_2}=\sum_{k_1=1}^{K_1} G_{k_1}  r_{k_1,k_2},
\eeq
$k_2=1,2,\ldots, K_2$.
Then the system is a $(\tilde P_{{\rm suc},1}(G), \tilde P_{{\rm suc},2}(G), \ldots, \tilde P_{{\rm suc},K_1}(G))$-Poisson receiver,
\beq{routing1111}
\tilde P_{{\rm suc}, k_1}(G)= \sum_{k_2=1}^{K_2} r_{k_1, k_2} P_{{\rm suc}, k_2}(\rho),
\eeq
$k_1=1,2,\ldots, K_1$.
\ethe

Using packet coding to increase system throughput has been widely addressed in the literature, see, e.g., \cite{casini2007contention,liva2011graph,paolini2012random,narayanan2012iterative,paolini2015coded,munari2015multi,stefanovic2018coded}.
The basic idea is to send a packet multiple times. If any one of them is successfully received, then the other copies of that packet can be removed from the system. Such an assumption is known as the {\em perfect} Successive Interference Cancellation (SIC) assumption.
The decoding process is known as a {\em peeling decoder} that repeatedly removes decoded packets from the system.
To analyze such a decoding process, the tree evaluation method \cite{luby1998analysis,luby1998analysisb,richardson2001capacity} tracks the evolution of the decodability probability after each SIC iteration. Such an analysis leads to the exact asymptotic system throughput under the {\em tree assumption}. The framework in \cite{chang2020Poisson} added the reduced Poisson offered load step in the  tree evaluation method
that leads to the following closure property for Poisson receivers with packet coding.

\bthe{coding}{\bf (Poisson receivers with packet coding \cite{chang2020Poisson})}
Consider a system with $G_{k} T$ class $k$ active users, $k=1,2, \ldots, K$, and $T$ independent
Poisson receivers with $K$ classes of input traffic and the success probability functions $P_{{\rm suc},1}(\rhog), P_{{\rm suc},2}(\rhog), \ldots, P_{{\rm suc},K}(\rhog)$. Each class $k$ user transmits its packet  for $L_k$ times (copies)  {\em uniformly} and {\em independently}
  to one of the $T$ Poisson receivers.
Let $\Lambda_{k,\ell}$ be the probability that a class $k$ packet is transmitted $\ell$ times, i.e.,
\begin{equation}
P(L_{k}=\ell)=\Lambda_{k,\ell}, \;\ell=1,2,\dots
\end{equation}
Define the function
\beq{mean0000mul}
\Lambda_{k}(x)=\sum_{\ell=0}^\infty \Lambda_{k,\ell} \cdot x^\ell ,
\eeq
and the function
\beq{mean3333mul}
\lambda_{k}(x)=\frac{\Lambda^\prime(x)}{\Lambda^\prime(1)}.
\eeq
Denote by $\odot$ the element-wise multiplication of two vectors, i.e., for two vectors
$u=(u_1, u_2, \ldots, u_K)$ and $v=(v_1, v_2, \ldots, v_K)$,
\beq{element1111}
u \odot v =(u_1 v_1, u_2 v_2, \ldots, u_K v_K).
\eeq
Then under the perfect SIC assumption and the tree assumption for a very large $T$, the system of coded Poisson receivers after the $i^{th}$ SIC iteration is a Poisson receiver with the success probability function for a tagged class $k$ packet
\beq{mean8888thumula}
 \tilde P_{{\rm suc},k}^{(i)}(G)=1-\Lambda_k \Big  (1- P_{{\rm suc},k}(q^{(i-1)} \odot G \odot \Lambda^\prime(1))\Big),
\eeq
$k=1,2, \ldots, K$, where $q^{(i)}=(q_{1}^{(i)}, q_{2}^{(i)}, \ldots, q_{K}^{(i)})$ can be computed recursively from the following equation:
\bear{tag6666dmul}
q_{k}^{(i+1)}&=&\lambda_{k}\Big (1- P_{{\rm suc},k}(q^{(i)} \odot  G \odot \Lambda^\prime(1))\Big ),\label{eq:tag6666bmul}
\eear
with $q^{(0)}=(1, 1, \ldots, 1)$.
\ethe

\bsubsec{Rayleigh  block fading channel with capture}{fading}

In this section, we consider the block fading channel with capture in \cite{stefanovic2014exploiting,clazzer2017irregular,stefanovic2018coded}.
We show that such a channel model can also be modelled by a Poisson receiver.
In a wireless channel with $N$ active users and one receiver,
the signal at the receiver is commonly represented by the sum of transmitted signals and the added white Gaussian noise, i.e.,
\beq{fading1111}
\sum_{n=1}^N h_n s_n +{\cal N},
\eeq
where $s_n$ is the signal transmitted by the $n^{th}$ active user, ${\cal N}$ is the added white Gaussian noise, and $h_n$ is the channel gain between the $n^{th}$ active user and the receiver.
Suppose that (i) the transmitted signals are orthogonal, (ii) the transmitted power at each active user is ${\cal P}$, and (iii) the noise power is ${\cal P}_{noise}$. Then
the total received power at the receiver is
\beq{fading1155}
\sum_{n=1}^N ||h_n||^2 {\cal P} +{\cal P}_{noise}.
\eeq
In the threshold-based decoding model, the signal $s_i$ can be successfully decoded if
the signal-to-interference-and-noise ratio (SINR) is
higher then a predefined threshold $b^\star$, i.e.,
\beq{fading2222}
\frac{||h_i||^2 {\cal P}}{\sum_{n \ne i} ||h_n||^2 {\cal P} +{\cal P}_{noise}} \ge b^\star.
\eeq
Let $X_n=||h_n||^2$, $\gamma={\cal P}/{\cal P}_{noise}$, and $b=b^\star/{\cal P}$. Then
\req{fading2222} can be written as follows:
\beq{fading3333}
\frac{X_i}{\sum_{n \ne i} X_n +\frac{1}{\gamma}} \ge b.
\eeq
As in \cite{stefanovic2014exploiting,clazzer2017irregular,stefanovic2018coded}, we assume the independent Rayleigh fading model for each active user,  i.e., $X_n$'s are independent and exponentially distributed with mean 1.
Then the probability that the signal $s_i$ can be successfully decoded  is
\beq{fading4444}
\pr (\frac{X_i }{\sum_{n \ne i} X_n  +\frac{1}{\gamma}} \ge b)=\frac{e^{-b/\gamma}}{(1+b)^{N-1}}.
\eeq

Now we consider the capture effect in the threshold-based model.
Like the SIC decoding algorithm described in the previous section,  the threshold-based model with the {\em capture} effect first decodes the signal with the largest power. If the SINR of the signal with the largest power is not smaller than the threshold, then that signal is successfully decoded, and it is removed from the received signal. The process is then repeated until no signal can be successfully decoded.
The probability that the $r$ signals $s_i$, $i=1,2,\ldots, r$, are successfully decoded in the order $1,2, \ldots, r$, is
\bear{fading5555}
&&\pr \Big (\frac{X_1 }{\sum_{n =2}^N X_n  +\frac{1}{\gamma}} \ge b,\frac{X_2 }{\sum_{n =3}^N X_n  +\frac{1}{\gamma}} \ge b,
\nonumber  \\
&&\quad \quad \ldots,\frac{X_{r} }{\sum_{n =r+1}^N X_n  +\frac{1}{\gamma}} \ge b\Big )\nonumber \\
&& =\frac{e^{-\frac{1}{\gamma}((1+b)^{r}-1)}}{(1+b)^{r\left(N-1-\frac{r-1}{2}\right)}}.
\eear
The detailed derivation of \req{fading5555} is shown in Appendix A.
As such, the probability that there are at least $r$ successfully decoded signals is
\bear{fading5566}
\frac{N!}{(N-r)!}\frac{e^{-\frac{1}{\gamma}((1+b)^{r}-1)}}{(1+b)^{r\left(N-1-\frac{r-1}{2}\right)}}.
\eear
Since $E[X]=\sum_{r=1}^\infty \pr (X \ge r)$ for any nonnegative discrete random variable $X$, we know that the average number of successfully decoded signals
is
\bear{fading6666}
\sum_{r=1}^N \frac{N!}{(N-r)!}\frac{e^{-\frac{1}{\gamma}((1+b)^{r}-1)}}{(1+b)^{r\left(N-1-\frac{r-1}{2}\right)}}.
\eear

For a Poisson offered load $\rho$, the number of active users $N$ follows a Poisson distribution with mean $\rho$.
From \req{fading6666}, the throughput for the Rayleigh  block fading channel with capture (subject to a Poisson offered load $\rho$) is
\bear{fading7777}
&&S=\sum_{N=0}^\infty \frac{e^{-\rho} \rho^N}{N!} \sum_{r=1}^N \frac{N!}{(N-r)!}\frac{e^{-\frac{1}{\gamma}((1+b)^{r}-1)}}{(1+b)^{r\left(N-1-\frac{r-1}{2}\right)}}\nonumber \\
&&=\sum_{N=0}^\infty  \sum_{r=1}^N \frac{e^{-\rho} \rho^N}{(N-r)!} \frac{e^{-\frac{1}{\gamma}((1+b)^{r}-1)}}{(1+b)^{r\left(N-1-\frac{r-1}{2}\right)}}.
\eear
Thus, the Rayleigh  block fading channel with capture is a $P_{\rm suc}(\rho)$-Poisson receiver with
\bear{fading8888}
P_{\rm suc}(\rho)&=&\sum_{N=0}^\infty  \sum_{r=1}^N \frac{e^{-\rho} \rho^{N-1}}{(N-r)!} \frac{e^{-\frac{1}{\gamma}((1+b)^{r}-1)}}{(1+b)^{r\left(N-1-\frac{r-1}{2}\right)}} \nonumber \\
&=&\sum_{N=0}^\infty  \sum_{\rone=0}^{N-1} \frac{e^{-\rho} \rho^{N-1}}{(N-(\rone+1))!} \frac{e^{-\frac{1}{\gamma}((1+b)^{\rone+1}-1)}}{(1+b)^{(\rone+1)\left(N-1-\frac{\rone}{2}\right)}} \nonumber \\
&=&\sum_{t=0}^\infty  \sum_{\rone=0}^{t} \frac{e^{-\rho} \rho^t}{(t-\rone)!} \frac{e^{-\frac{1}{\gamma}((1+b)^{\rone+1}-1)}}{(1+b)^{(\rone+1)\left(t-\frac{\rone}{2}\right)}}.
\eear

\bsec{ALOHA Receivers}{phiALOHA}

The framework of Poisson receivers in \cite{chang2020Poisson} is a {\em probabilistic} framework for analyzing coded random access.
 Such a framework relies on the tree assumption to keep the {\em independence} of various classes of the input traffic. For the tree assumption to hold, the number of Poisson receivers $T$ in \rthe{coding} needs to be very large. To deal with the scenario where input traffic independence is difficult to justify, we propose a deterministic framework, called {\em ALOHA} receivers.

 \bsubsec{Definitions and examples of ALOHA receivers}{definALOHA}

 Denote by ${\cal Z}^+$  the set of nonnegative integers. We say a system with $K$ classes of input traffic is subject to a {\em deterministic} load   $n=(n_1, n_2, \ldots, n_K)  \in {{\cal Z}^+}^K$ if the number of class $k$ packets arriving at
the system is $n_k$. For two vectors  $n^{\prime}$ and $n^{\prime\prime}$ in ${{\cal Z}^+}^K$, we say $n^{\prime} \le n^{\prime\prime}$ if $n^{\prime}_k \le  n^{\prime\prime}_k$ for all $k=1,2,\ldots, K$.

\begin{figure}[ht]
	\centering
	\includegraphics[width=0.45\textwidth]{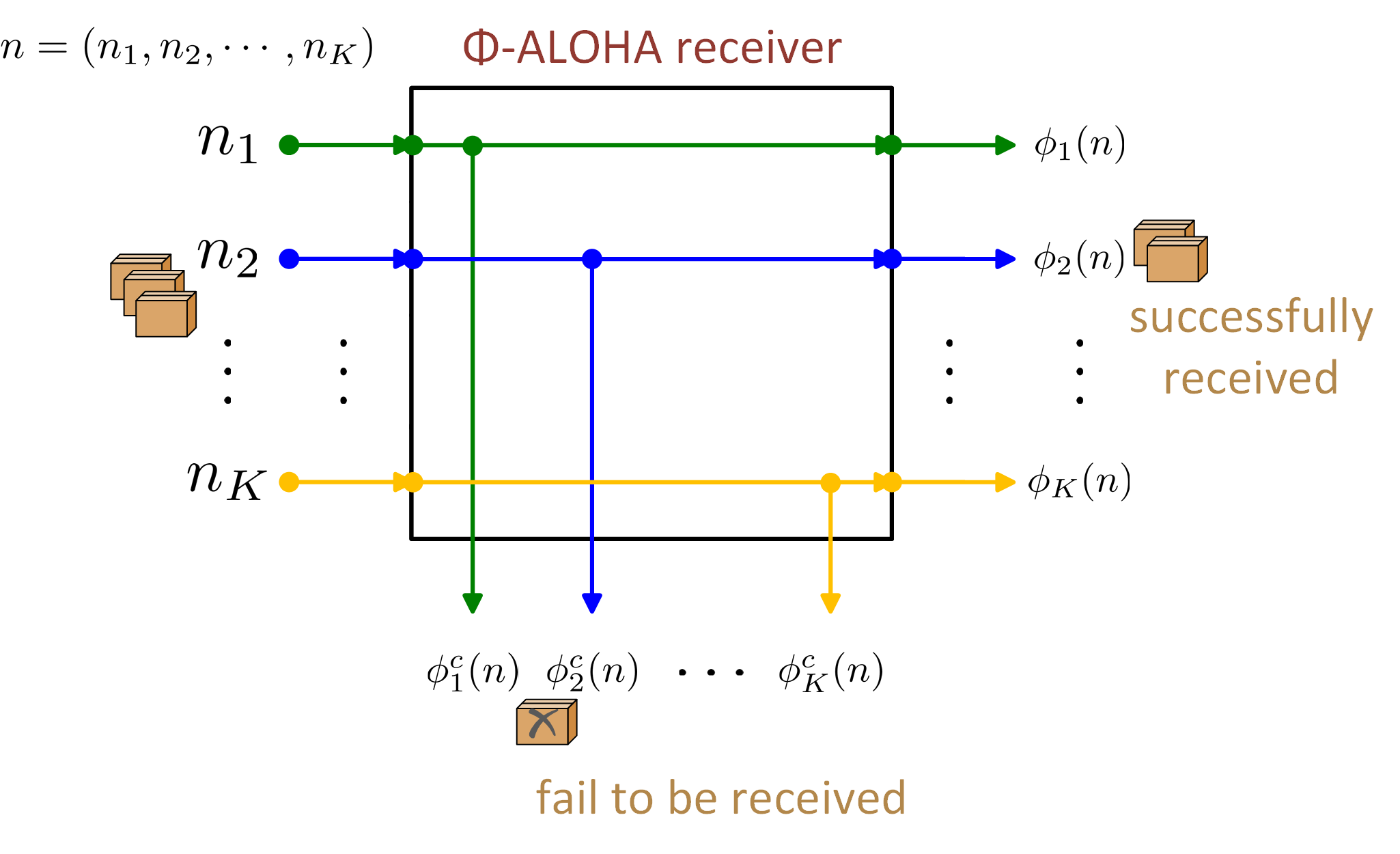}
	\caption{A $\phi$-ALOHA receiver with $K$ classes of input traffic.}
	\label{fig:phiALOHA}
\end{figure}

 \bdefin{phiALOHA}{\bf (ALOHA receiver with multiple classes of input traffic)}
 Consider a  deterministic function
 $$\phi: {{\cal Z}^+}^K \rightarrow {{\cal Z}^+}^K$$ that
maps a $K$-vector $n=(n_1, n_2, \ldots, n_K)$ to the $K$-vector
$(\phi_1(n), \phi_2(n), \ldots, \phi_K(n))$.
 An abstract receiver is called $\phi$-ALOHA receiver (with $K$ classes of input traffic) if  the number of class $k$ packets that are successfully received is exactly $\phi_k(n)$, $k=1,2, \ldots, K$, when the receiver is subject to a deterministic load $n=(n_1, n_2, \ldots, n_K)$ (see \rfig{phiALOHA}).
 The function $\phi$ is called the {\em success} function of the ALOHA receiver. Define the {\em failure} function $\phi^c$
 by
 \beq{phi0012}
 \phi^c(n)= n-\phi(n).
 \eeq
 A  $\phi$-ALOHA receiver is called {\em monotone} if the {\em failure} function $\phi^c$
  is {\em increasing} in the deterministic load $n$, i.e.,
for any $n^{\prime} \le n^{\prime\prime}$,
\beq{phi0000}
\phi^c(n^{\prime}) \le \phi^c(n^{\prime\prime}).
\eeq
\edefin

The naming of ALOHA receivers is from the Slotted ALOHA (SA) system \cite{ALOHA}.
In the classical collision channel model, if there is more than one packet transmitted to a receiver, then there is a collision, and collided packets are assumed to be lost. On the other hand, if there is exactly one packet transmitted to a receiver, then that packet is assumed to be successfully received. Thus, the SA system is a $\phi$-ALOHA receiver with a single class of input traffic, where
\beq{phi7711}
\phi(n)=\left \{\begin{array}{cc}
		1 & \mbox{if}\; n =1 \\
		0 & \mbox{otherwise}
	\end{array} \right ..
\eeq
The $D$-fold ALOHA system proposed in \cite{ordentlich2017low} is a generalization of the SA system.
If there are less than or equal to  $D$ packets transmitted in a time slot, then all these packets
can be successfully decoded. On the other hand, if there are more than $D$ packets
transmitted in a time slot, then all these packets are lost.
Clearly, the $D$-fold ALOHA system in a time slot is a monotone $\phi$-ALOHA receiver with a single class of input traffic, where
\beq{phi7722}\phi(n)=\left \{\begin{array}{cc}
		n & \mbox{if}\; n \le D \\
		0 & \mbox{otherwise}
	\end{array} \right ..
\eeq
In this paper, we simply call the $D$-fold ALOHA system in a time slot a $D$-fold ALOHA receiver.

In addition to the SA system with a single class of input traffic, one can use the $\phi$-ALOHA receiver to model the near-far SIC decoding
in the following example.

\bex{nearfar}{(Near-far SIC decoding)}
 Suppose that there are two classes of input traffic to a receiver.
The power of a class 1 packet at the receiver is much stronger than that of a class 2 packet at the receiver.
One may view class 1 (resp. 2) packets as the users who are near (resp. far from) the receiver. Suppose that there are two packets arriving at the receiver: one is a class 1 packet, and the other is a class 2 packet. The SIC decoding algorithm first decodes the class 1 packet  and then removes it to reduce the interference to the class 2 packet (under the perfect SIC assumption). By doing so, the class 2 packet can also be decoded.
For such a near-far SIC decoding, we can model it as a monotone $\phi$-ALOHA receiver with
$$\phi(n_1,n_2)=\left \{\begin{array}{cc}
		(n_1,n_2) & \mbox{if}\; (n_1,n_2) \le (1,1) \\
		(0,0) & \mbox{otheriwse}
	\end{array} \right .. $$
\eex

Note that
a $\phi$-ALOHA receiver with $K_1$ classes of input traffic and
a $\psi$-ALOHA receiver with $K_2$ classes of input traffic in parallel can be viewed as a $(\phi,\psi)$-ALOHA receiver with $K_1+K_2$ classes of input traffic. The near-far SIC decoding model in \rex{nearfar} can be viewed as two SA systems in parallel, where the first SA system is for class 1 traffic, and the second SA system is for class 2 traffic.

In the following theorem, we show that a $\phi$-ALOHA receiver is also a Poisson receiver. Such a Poisson receiver is called the induced Poisson receiver from the $\phi$-ALOHA receiver. This allows us to incorporate our deterministic framework of ALOHA receivers into the probabilistic framework of Poisson receivers.

\bthe{phimain}
A $\phi$-ALOHA receiver with $K$ classes of input traffic is  a $(P_{{\rm suc},1}(\rhog), P_{{\rm suc},2}(\rhog), \ldots, P_{{\rm suc},K}(\rhog))$-Poisson receiver, where
\beq{phi2222}
P_{{\rm suc},k}(\rhog)=\frac{1}{\rho_k}\sum_{n}\phi_k(n)\prod_{\ell=1}^K \frac{e^{-\rho_\ell} {\rho_\ell}^{n_\ell}}{{n_\ell}!},
\eeq
$k=1,2, \ldots, K$.
\ethe

\bproof
Note that the throughput of class $k$ packets in a $\phi$-ALOHA receiver subject to a Poisson offered load $\rho=(\rho_1, \ldots, \rho_K)$ is
\beq{phi1111}
S_k=\sum_{n}\phi_k(n)\prod_{\ell=1}^K \frac{e^{-\rho_\ell} {\rho_\ell}^{n_\ell}}{{n_\ell}!}.
\eeq
Using \req{Poithrmul} yields \req{phi2222}.


\eproof

%
\bsubsec{Operations on functions}{algebra}

For the ease of our presentation, we introduce several operations on functions and their algebraic properties.
Consider the class of functions
\beq{alg0077}
{\cal F}=\{f: {{\cal Z}^+}^K \rightarrow {{\cal Z}^+}^K\;\mbox{with}\; f(n) \le n\}.
\eeq
Define two binary operations on this class of functions: the minimum operation $\wedge$ and the composition operation $\circ$.
\bear{alge0000}
(f \wedge g)(n) &=&\min[f(n), g(n)], \label{eq:alge0000a} \\
(f \circ g)(n)&=& f(g(n)). \label{eq:alge0000b}
\eear
The minimum operation in \req{alge0000a} is a component-wise operation.
We say $f = (\mbox{resp.}\;\le) g$ if $f(n) =(\mbox{resp.}\;\le)  g(n)$ for all $n \in {{\cal Z}^+}^K$.

It is easy to see that
these two operations have the following properties:
\bear{alge0022}
f \wedge g &=&g \wedge f, \quad\mbox(commutativity)\\
f \wedge (g \wedge h)&=&(f \wedge g) \wedge h, \quad\mbox(associativity)\\
f \circ (g \circ h)&=& (f \circ g) \circ h. \quad\mbox(associativity)
\eear
Let $\epsilon$ be the identity mapping, i.e., $\epsilon(n)=n$. Clearly,
$\epsilon$ is the identity element for these two operations, i.e.,
\bear{alge0025}
&&f \wedge \epsilon =\epsilon \wedge f=f, \\
&&f \circ \epsilon =\epsilon \circ f =f.
\eear

A function is said to be increasing if
\beq{alge0055}
f(n^{\prime}) \le f(n^{\prime\prime}),
\eeq
for all $n^{\prime} \le n^{\prime\prime}$.
Define the class of increasing  functions
\beq{alge1111}
\finc=\{f\in {\cal F}:f\; \mbox{is increasing}.\}.
\eeq

In the following proposition, we state the closure property and the monotone property for the two operations in $\finc$. The proofs are rather straightforward and thus omitted.
\bprop{monotone}
\begin{description}
\item[(i)] (Closure property) If $f,g$ are in $\finc$, then $f \wedge g$ and $f\circ g$ are also in $\finc$.
\item[(ii)] (Monotone property) For $f_1, f_2, g_1, g_2$ in $\finc$, if
$f_1 \le f_2$, $g_1 \le g_2$,  then
\bear{alge0066}
&&f_1 \wedge g_1 \le f_2 \wedge g_2,\\
&&f_1 \circ g_1 \le f_2 \circ g_2.
\eear
\end{description}
\eprop

Let $f^{(0)}=\epsilon$ and
$f^{(i+1)}=f^{(i)} \circ f$.
For any $f \in \finc$, we have
\beq{alge2222}
f^{(i+1)}=  f^{(i)} \circ f \le f^{(i)} \circ \epsilon  = f^{(i)}.
\eeq
As such, $\{f^{(i)}(n), i=1,2,\ldots\}$ is a decreasing sequence and thus converges to a vector $n^*$.
In view of this, we can define a unary operation  $*$, called the closure operation, on a function $f$ as the limit of the decreasing sequence of functions $\{f^{(i)}, i=1,2,\ldots\}$, i.e.,
\beq{alge3333}
f^*=\lim_{i \to \infty} f^{(i)}.
\eeq

\bex{starshape}{(Star-shaped functions)}
For $K=1$, a function $f$ is called
 a {\em star-shaped} function if $f(n)/n$ is increasing in $n$ for all $n\ge 1$.
 As $f \in \finc$, $f(n)/n \le 1$. Let
 $$n_1=\inf\{n: f(n)/n=1\}.$$
  Then for $n \ge n_1$,
 we have $f(n)/n \ge f(n_1)/n_1=1$ and thus $f(n)=n$ for $n \ge n_1$.
On the other hand, for $n <n_1$, we have $f(n)/n< 1$.
If $f^{(i)}(n) \ge 1$, then for $n <n_1$,
 \bearn
\frac{f^{(i+1)}(n)}{f^{(i)}(n)} =\frac{f(f^{(i)}(n))}{f^{(i)}(n)} \le \frac{f(n)}{n}<1.
\eearn
This implies that $f^{(i+1)}(n) <f^{(i)}(n)$. Thus, $\{f^{(i)}(n), i=1,2,\ldots\}$ decreases to 0 for $n <n_1$.
As such, we have
\beq{stars1111}
f^*(n)=\left \{\begin{array}{cc}
		0 & \mbox{if}\; n < n_1 \\
		n & \mbox{otherwise}
	\end{array} \right ..
\eeq
\eex

In the following lemma, we show several properties of the closure operation.
\blem{closure}
For any $f,g$ in $\finc$, we have
\begin{description}
\item[(i)] (Monotone property) If $f \le g$, then $f^* \le g^*$.
\item[(ii)] $f \circ f^*= f^* \circ f=f^*$.
\item[(iii)] $f^* \circ f^*= f^*$.
\item[(iv)] $(f^*)^* =f^*$.
\item[(v)]
$(f \circ g)^* =(g \circ f)^*$
\item[(vi)]
$(f \circ g)^*=(f^* \circ g^*)^*$.
\item[(vii)]
$(f \circ g)^*=(f \wedge g)^*$.
\end{description}
\elem

\bproof
(i) From the monotone property in \rprop{monotone}, we have $f^{(i)} \le g^{(i)}$ for all $i$.
Letting $i \to \infty$ completes the argument.

(ii) That (ii) holds  follows trivially from the definition of the closure operation as the limit of the decreasing sequence of functions $\{f^{(i)}, i=0,1,2,\ldots\}$.

(iii) From (ii), it follows that
$f^* \circ f^{(i)} =f^*$ for all $i$. Letting $i \to \infty$ completes the argument.

(iv) From (iii), we have $(f^*)^{(i)}=f^*$ for all $i$. Letting $i \to \infty$ completes the argument.

(v) First, note from the associative property of $\circ$ that
\beq{alge5555}
(f \circ g)^{(i)} \circ f= f \circ (g \circ f)^{(i)}.
\eeq
Letting $i \to \infty$ yields
\beq{alge6666}
(f \circ g)^{*} \circ f= f \circ (g \circ f)^{*}.
\eeq
Since $f,g \le \epsilon$, we then have the monotone property in \rprop{monotone} that
\bear{alge7777}
&&(f \circ g)^{*} = (f \circ g)^{*} \circ \epsilon \ge (f \circ g)^{*} \circ f\nonumber \\
&&= f \circ (g \circ f)^{*} =(\epsilon \circ f) \circ (g \circ f)^{*} \nonumber\\
&&\ge (g \circ f) \circ (g \circ f)^{*}=(g \circ f)^{*}.
\eear
Interchanging $f$ and $g$ in \req{alge7777} yields
\beq{alge7788}
(g \circ f)^{*}\ge (f \circ g)^{*}.
\eeq
Thus, we have from \req{alge7777} and \req{alge7788} that
\beq{alge7799}
(g \circ f)^{*}=(f \circ g)^{*}.
\eeq

(vi) Since $f \ge f^*$ and $g \ge g^*$, we have from the monotone property in (i) of this lemma that
$$(f \circ g)^*\ge (f^* \circ g^*)^*. $$
On the other hand, note that $f \ge (f \circ g)$ and $g \ge (f \circ g)$.
It then follows from the monotone property that $f^* \ge (f \circ g)^*$ and $g^* \ge (f \circ g)^*$. Thus,
from \rlem{closure}(iii),
$$(f \circ g)^*=(f \circ g)^* \circ (f \circ g)^* \le f^* \circ g^*.$$
Using (iv) of this lemma and the monotone property in (i) of this lemma yields
$$(f \circ g)^*=((f \circ g)^*)^* \le (f^* \circ g^*)^*.$$

(vii) Since $f \ge f \circ g$ and $g \ge f \circ g$,
we have from \rprop{monotone}(ii) that
$$f \wedge g \ge (f \circ g) \wedge (f \circ g)=f \circ g.$$
It then follows from the monotone property in (i) of this lemma that
$$(f \wedge g)^* \ge (f \circ g)^*.$$
On the other hand, we also have from $f \wedge g \le f$ and $f \wedge g \le g$ that
$$(f \wedge g)^{(2)}= (f \wedge g) \circ (f \wedge g) \le f \circ g.$$
This then leads to
$$(f \wedge g)^* \le (f \circ g)^*.$$
\eproof

In addition to the closure operation, we also need another unary operation $c$, call the {\em complement} operation of a function $f \in {\cal F}$. Specifically, we denote by $f^c$ the complement function of $f$ with
\beq{alge8888}
(f^c)(n)= n-f(n).
\eeq
Clearly, $f^c$ is in ${\cal F}$ and $(f^c)^c=f$. In our definition of ALOHA receivers, the success function $\phi$ and the failure function $\phi^c$ are complement functions of each other. Moreover, if a $\phi$-ALOHA receiver is {\em monotone}, then the failure function $\phi^c$ is in $\finc$. As such, we can use the closure operation on the failure function $\phi^c$.

\bsubsec{Two ALOHA receivers in tandem}{tandem}

Consider a system
with two   receivers  and $K$ classes of input traffic. The first receiver is a $\phi$-ALOHA receiver, and the second receiver is a $\psi$-ALOHA receiver. These two receivers are subject to the same $K$ classes of input traffic.
The two receivers are arranged {\em in tandem} so that
only packets that are successfully received by receiver 1 can be forwarded to receiver 2 for SIC decoding, but not the other way around.
For a deterministic load $n=(n_1, n_2, \ldots,  n_K)$,
there are $\phi_k(n)$ class $k$ packets that are successfully received by receiver 1, and thus
the number of class $k$ packets arriving at receiver 2 is effectively reduced from $n_k$ to $n_k-\phi_k(n)$, $k=1,2, \ldots, K$.
As such, receiver 2 is subject to a deterministic load $n-\phi(n)$. Since receiver 2 is a $\psi$-ALOHA receiver, the number of class $k$ packets that are successfully received by receiver 2 is
$\psi_k(n-\phi(n))$. Thus, the total number of class $k$ packets that are successfully received by
these two receivers is
$$\phi_k(n)+\psi_k\Big (n-\phi(n)\Big ).$$
This shows that the system is a $\zeta$-ALOHA receiver, where
\beq{tandem1111b}
\zeta_k(n)=\phi_k(n)+\psi_k\Big (n-\phi(n)\Big ),
\eeq
$k=1,2, \ldots, K$.

Another way to see this is to count the numbers of packets that are not successfully received.
After the receiver 1, the deterministic load is effectively reduced from $n$ to $\phi^c(n)$. Then
after receiver 2, the deterministic load is further reduced from  $\phi^c(n)$ to $\psi^c(\phi^c(n))=(\psi^c \circ \phi^c)(n)$.  Thus, the numbers of packets that are successfully received are $(\psi^c \circ \phi^c)^c(n)$.
This is stated in the following theorem.

\begin{figure}[ht]
	\centering
	\includegraphics[width=0.4\textwidth]{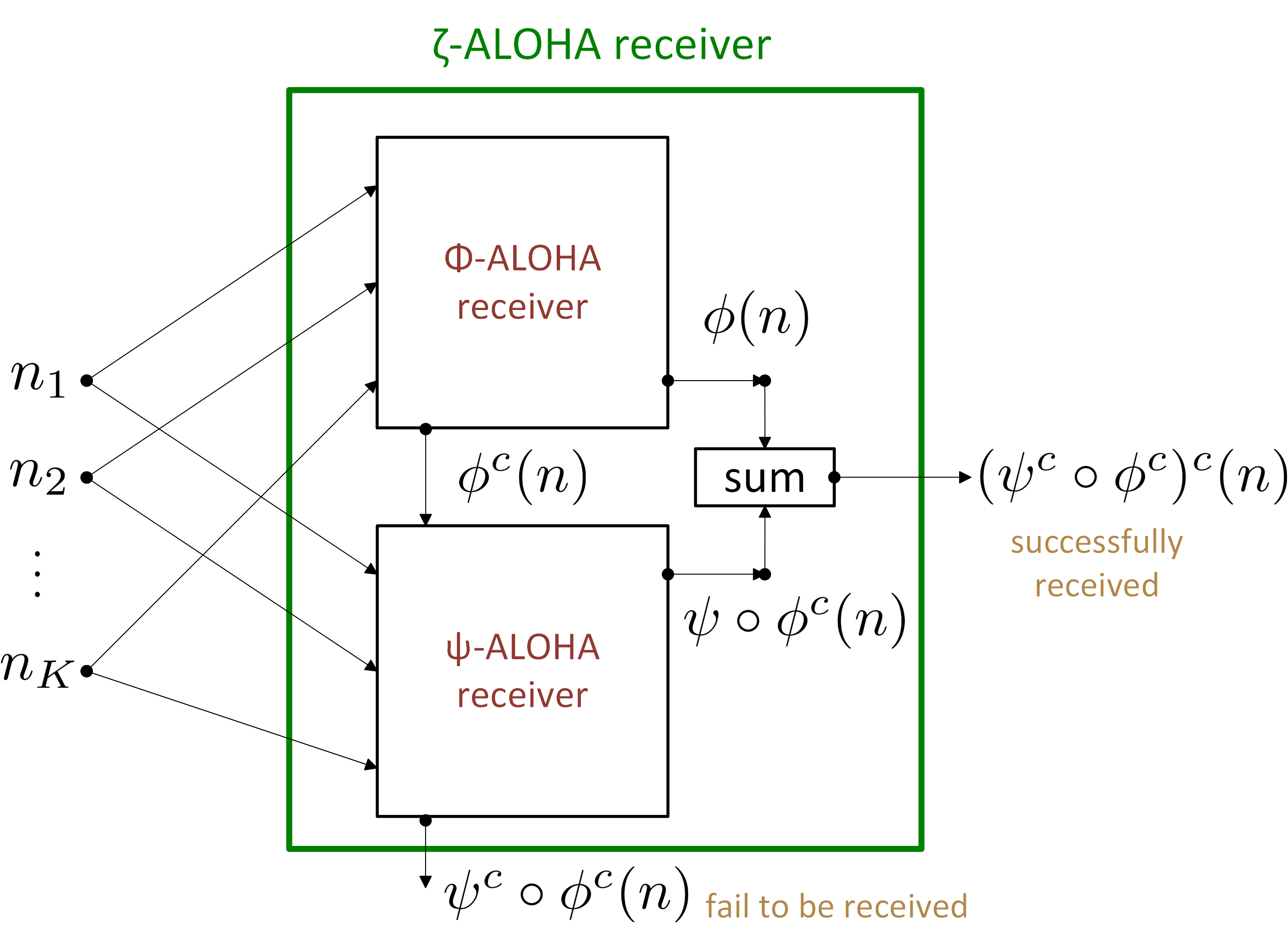}
	\caption{A system with a $\phi$-ALOHA receiver and  a $\psi$-ALOHA receiver in tandem.}
	\label{fig:tandem}
\end{figure}

\bthe{tandem} For the system
with a $\phi$-ALOHA receiver and  a $\psi$-ALOHA receiver in tandem (see \rfig{tandem}), it is a $\zeta$-ALOHA receiver,
where
\beq{tandem8888}
\zeta=(\psi^c \circ \phi^c)^c .
\eeq
\ethe


\bsubsec{Two cooperative ALOHA receivers}{coop}

Consider the same setting as in  \rsec{tandem}. Now we assume that these two receivers are cooperative and
packets that are successfully received by receiver 2 can also be forwarded to receiver 1 for SIC decoding.
As such, we can repeat the SIC decoding between these two receivers until no packets can be decoded.
To illustrate this,
let $n^{(i)}=(n_1^{(i)}, n_2^{(i)}, \ldots, n_K^{(i)})$, where $n_k^{(i)}$ is the number of class $k$ packets remained to be decoded in the system after the $i^{th}$ SIC iteration, $k=1,2,\ldots, K$,  and $i=0,1,\ldots$. Clearly, $n^{(0)}=n$.
Note that the first SIC iteration corresponds to the two receivers in tandem in  \rsec{tandem}.
After the first SIC iteration, the number of class $k$ packets that are remained to be decoded is
\beq{tandem1111c}
n_k^{(1)}=n_k^{(0)}-\zeta_k(n^{(0)})
\eeq
where $\zeta_k$ is defined in \req{tandem1111b}.
In general, we have
\bear{tandem2233}
n^{(i+1)}&=&n^{(i)}-\zeta(n^{(i)})\nonumber\\
&=&\zeta^c(n^{(i)})=(\zeta^c)^{(i)}(n).
\eear
As this sequence of vectors $\{n^{(i)}, i \ge 0\}$ is nonnegative and monotonically decreasing, it converges
to the vector
\beq{tandem2266}
n^{(\infty)}=(\zeta^c)^{*}(n)=(\psi^c \circ \phi^c)^*(n).
\eeq
This leads to the following theorem.

\begin{figure}[ht]
	\centering
	\includegraphics[width=0.35\textwidth]{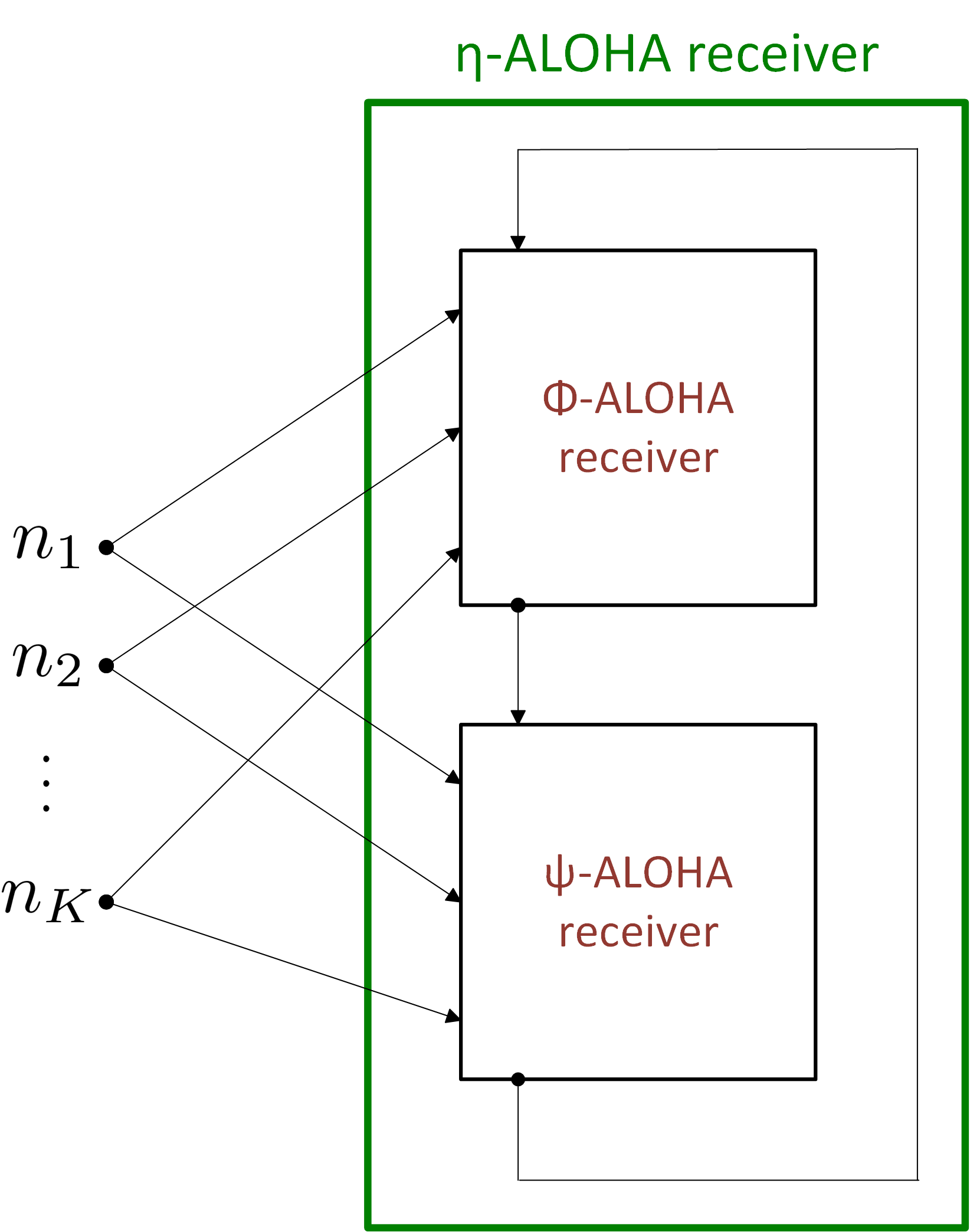}
	\caption{Two cooperative receivers: a $\phi$-ALOHA receiver and  a $\psi$-ALOHA receiver.}
	\label{fig:coop}
\end{figure}

\bthe{coop} Consider a system
with a $\phi$-ALOHA receiver and  a $\psi$-ALOHA receiver (see \rfig{coop}). If these two receivers are cooperative, then  it is a $\eta$-ALOHA receiver,
where
\beq{tandem2244}
\eta=((\psi^c \circ \phi^c)^*)^c .
\eeq
\ethe

One interesting question is whether the system converges to the same $\eta$-ALOHA receiver in \req{tandem2244} if the  SIC decoding order of the two receivers is interchanged.
In the following theorem, we show this is true if the two receivers are monotone.

\bthe{interchange}
Suppose that both the  $\phi$-ALOHA receiver and the $\psi$-ALOHA receivers are {\em monotone}.
Then the system with two cooperative receivers converges to the same $\eta$-ALOHA receiver in \req{tandem2244} if the SIC decoding order of the two receivers is interchanged, i.e., the order of the SIC decoding process does not affect the decoding result.
Moreover, the system with two cooperative receivers is also {\em monotone}.
\ethe

\bproof
Note that if we interchange the decoding order of the two receivers, then it is a $((\phi^c \circ \psi^c)^*)^c$-ALOHA receiver (from \rthe{coop}). Since both the $\phi$-ALOHA receiver and the $\psi$-ALOHA receiver are monotone,
both $\phi^c$ and $\psi^c$ are in $\finc$. As a result of \rlem{closure} (v), we have
$$(\psi^c \circ \phi^c)^*=(\phi^c \circ \psi^c)^*.$$
Thus, the system converges to the same $\eta$-ALOHA receiver in \req{tandem2244} if the  SIC decoding order of the two receivers is interchanged.

Since $(\psi^c \circ \phi^c)^*$ is still in $\finc$ from the closure property in \rprop{monotone}(i), the $\eta$-ALOHA receiver in \req{tandem2244} is also monotone.
\eproof

By using an inductive argument, we note that the result in \rthe{interchange} still holds for the scenario with more than two monotone cooperative receivers.

\bsubsec{ALOHA receivers with traffic multiplexing}{multiplex}


In this section, we consider ALOHA receivers with traffic multiplexing. To provide insight into our analysis, we
first consider a system with $K$ classes of (external) input traffic that are multiplexed into a
$D$-fold ALOHA receiver. Recall that  a $D$-fold ALOHA receiver is a $\phi$-ALOHA receiver with $\phi$ in \req{phi7722}.
Let $n=(n_1, n_2,\ldots, n_K)$ be the deterministic load to the system.
As the $K$ classes of (external) input traffic are multiplexed into the
$D$-fold ALOHA receiver, the deterministic load to the $D$-fold ALOHA receiver is $\sum_{i=1}^K n_i$. Thus, the number of class $k$ packets that are successfully received is $n_k$  if $\sum_{i=1}^K n_i \le D$. On the other hand, if $\sum_{i=1}^K n_i  > D$, then none of the arriving packets can be successfully received. This shows that such a system is a $\psi$-ALOHA receiver with the success function
\beq{multiplex1111}
\psi(n)=\left \{\begin{array}{cc}
		(n_1,n_2,\ldots, n_K) & \mbox{if}\; \sum_{i=1}^K n_i \le D \\
		(0,0,\ldots,0) & \mbox{otherwise}
	\end{array} \right ..
\eeq

One important insight of the above $D$-fold ALOHA receiver is that  the arriving packets are either all successfully received or all failed to be decoded.
This motivates us to introduce the notion of {\em all-or-nothing} ALOHA receivers.

\bdefin{allornothing} A $\phi$-ALOHA receiver with $K$ classes of input traffic is said to be an {\em all-or-nothing} receiver
if it satisfies the following two properties:
\begin{description}
\item[(i)] (All-or-nothing property) For all $n=(n_1, n_2, \ldots,n_K)$ and $k=1,2,\ldots, K$, either $\phi_k(n)=n_k$ or $\phi_k(n)=0$.
\item[(ii)] (On-off property) If  $n^{\prime} \ge n^{\prime\prime}$ and $\phi_k(n^{\prime})=n^{\prime}_k$ for some $k$, then $\phi_k(n^{\prime\prime})=n^{\prime\prime}_k$. On the other hand, if  $n^{\prime} \le n^{\prime\prime}$ and $\phi_k(n^{\prime})=0$ for some $k$, then $\phi_k(n^{\prime\prime})=0$.
\end{description}
\edefin

The on-off property in \rdef{allornothing} implies that an all-or-nothing $\phi$-ALOHA receiver is {\em monotone} as $\phi_k^c(n^{\prime}) \le \phi_k^c(n^{\prime\prime})$ $k=1,2, \ldots, K$, for all $n^{\prime} \le n^{\prime\prime}$.

Clearly, a $D$-fold ALOHA receiver  with a single class of input traffic is an all-or-nothing $\phi$-ALOHA receiver, where the success function $\phi$  is specified in \req{phi7722}. When it is subject to $K$ classes of external input traffic, it is an all-or-nothing $\psi$-ALOHA receiver with the success function $\psi$ in \req{multiplex1111}.

Traffic multiplexing that maps $K$  classes of external input traffic into $T$ classes of internal input traffic can be represented by a $K \times T$ bipartite graph,
where the $K$ classes of external input traffic are the $K$ nodes on the left (the external nodes) and the $T$ classes of internal input traffic are the $T$ nodes on the right (the internal nodes). The constraint for traffic multiplexing is that each external node has at most one edge.

Let $H=(H_{k,t})$ be the $K\times T$ bi-adjacency matrix for the bipartite graph, where $H_{k,t}=1$ if there is an edge
between the external node $k$ and the internal node $t$, and 0 otherwise. Denote by the set $C_t$  the set of nonzero elements in the $t^{th}$ column of $H$. As each external node has at most one edge, each  row of $H$  has at most one nonzero element, and thus the sets $C_t$, $t=1,2, \ldots, T$, are disjoint. Such a bi-adjacency matrix is called a {\em traffic multiplexing matrix} in this paper. Specifically, a traffic multiplexing matrix $H$ is a binary matrix, where each row of $H$ has at most one nonzero element.

\begin{figure}[ht]
	\centering
	\includegraphics[width=0.43\textwidth]{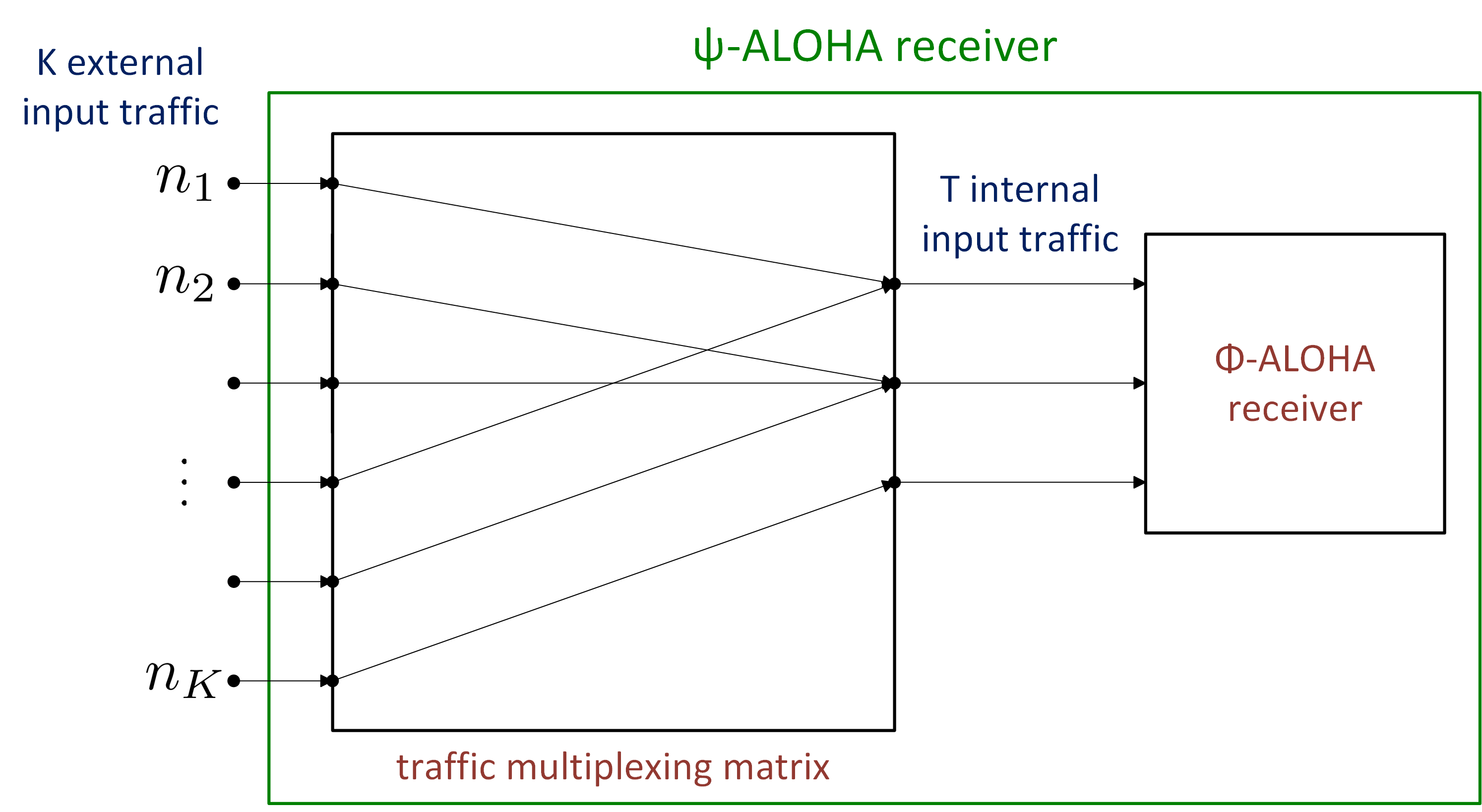}
	\caption{ALOHA receivers with traffic multiplexing: a system with $K$ classes of external  input traffic to
an all-or-nothing $\phi$-ALOHA receiver with $T$ classes of internal  input traffic through a $K \times T$ traffic multiplexing matrix $H$.}
	\label{fig:multiplexing}
\end{figure}

\bthe{multiplex} Consider a system with $K$ classes of external  input traffic to
an all-or-nothing $\phi$-ALOHA receiver with $T$ classes of internal  input traffic through a $K \times T$ traffic multiplexing matrix $H$ (see \rfig{multiplexing}). Then  the system is an all-or-nothing $\psi$-ALOHA receiver,
where
\beq{multiplex2222}
\psi_k(n)=\left \{\begin{array}{cc}
     n_k & \mbox{if}\;\sum_{t=1}^T H_{k,t} \cdot \phi_t (n H)>0\\
     0 & \mbox{otherwise}
	\end{array} \right ..
\eeq
$k=1,2, \ldots, K$.
\ethe

\bproof
Consider a deterministic load $n=(n_1, n_2, \ldots, n_K)$ for the external input traffic. Through traffic multiplexing, the load to the $\phi$-ALOHA receiver is $n H$.
As such, the number of internal class $t$ packets that are successfully received is  $\phi_t(n H)$, $t=1,2,\ldots, T$.
Since the $\phi$-ALOHA receiver is assumed to be an all-or-nothing receiver, the number of external class $k$ packets that are successfully received is $n_k$ if there is one internal class $t$ such that $H_{k,t}=1$ and the internal class $t$ packets are all successfully received, i.e., $\phi_t (n H)>0$.
Note that for each external class $k$, there is at most one internal class $t$ such that $H_{k,t}=1$. These two conditions can be simplified as the condition $\sum_{t=1}^T H_{k,t} \phi_t (n H)>0$.
This shows that the  all-or-nothing property.

Note that if an external class $k$ is not connected to any internal class, i.e., $H_{k,t}=0$ for all $t=1,2, \ldots, T$,
then none of class $k$ packets can be successfully received.
To show the on-off property, it thus suffices to consider an external class $k$ that is connected to an internal class $t(k)$.
In this case, we have from \req{multiplex2222} that
\beq{multiplex2222b}
\psi_k^c(n)=\left \{\begin{array}{cc}
     0 & \mbox{if}\; H_{k,t(k)} \cdot \phi_{t(k)}^c (n H)=0\\
     n_k & \mbox{otherwise}
	\end{array} \right ..
\eeq
Since an all-or-nothing receiver is monotone, we have  $\phi_{t(k)}^c(n^{\prime}H) \le \phi_{t(k)}^c(n^{\prime\prime}H)$, $k=1,2, \ldots, K$, for all $n^{\prime} \le n^{\prime\prime}$. Using this in \req{multiplex2222b} yields $\psi_k^c(n^{\prime}) \le \psi_k^c(n^{\prime\prime})$, $k=1,2, \ldots, K$. It is easy to see from the all-or-nothing property and the monotone property that the on-off property is also satisfied.
\eproof


\bsubsec{ALOHA receivers with packet coding}{codingALOHA}

Analogous to the iterative decoding algorithm for Poisson receivers with packet coding, we develop an iterative decoding algorithm for ALOHA receivers with packet coding.
As in traffic multiplexing in \rsec{multiplex}, we consider a system with $K$ classes of external input traffic. Inside the system, there is an
all-or-nothing $\phi$-ALOHA receiver with $T$ classes of internal input traffic.
A {\em packet coding} scheme for the $K$ classes of external input traffic is to {\em multicast} an external class $k$ packet to a set of internal classes, $B_k$, of the $\phi$-ALOHA receiver so that the sets $B_k$, $k=1,2, \ldots, K$, are disjoint. As traffic multiplexing in \rsec{multiplex}, a packet coding scheme also corresponds to a $K \times T$ bipartite graph with a
$K\times T$ bi-adjacency matrix $H$. The constraint for packet coding is that each {\em internal} node has at most one edge. Thus,
a packet coding matrix $H$ is a binary matrix, where each {\em column} of $H$ has at most one nonzero element.
Note that the set $B_k$ is the set of nonzero elements in the $k^{th}$ row of $H$.


Consider a deterministic load $n=(n_1, n_2, \ldots, n_K)$ for the external input traffic. As in \rsec{coop}, let $n^{(i)}=(n_1^{(i)}, n_2^{(i)}, \ldots, n_K^{(i)})$, where $n_k^{(i)}$ is the number of class $k$ packets remained to be decoded in the system after the $i^{th}$ SIC iteration, $k=1,2,\ldots, K$,  and $i=0,1,\ldots$. Clearly, $n^{(0)}=n$.
For the first SIC iteration, the load to the all-or-nothing $\phi$-ALOHA receiver is the $T$-vector $n^{(0)} H$.
As such, the number of internal class $t$ packets that are successfully received is  $\phi_t(n^{(0)} H)$, $t=1,2,\ldots, T$.
Since an external packet is successfully received if one of its multicast copies is successfully received,
the number of external class $k$ packets that are successfully received is $\max_{t \in B_k}\phi_t(n^{(0)} H)$, $k=1,2, \ldots, K$ (as the $\phi$-ALOHA receiver is an all-or-nothing receiver).
After the first iteration, the number of external class $k$ packets that are remained to be decoded is
\beq{codingA1111}
n^{(1)}_k= n^{(0)}_k- \max_{t \in B_k}\phi_t(n^{(0)} H),
\eeq
 $k=1,2, \ldots, K$. As the $\phi$-ALOHA receiver is an all-or-nothing receiver, we know that
 $n^{(1)}_k$ is either 0 or $n^{(0)}_k$. Such an all-or-nothing property is preserved through every SIC iteration.
 Define the function $\theta(n)=(\theta_1(n),\ldots, \theta_K(n))$ with
\beq{codingA2222}
 \theta_k(n)=\max_{t \in B_k}\phi_t(n H).
 \eeq
 Then
  we have from the argument for \req{codingA1111} that
\beq{tandem2233c}
n^{(i+1)}=n^{(i)}-\theta(n^{(i)}).
\eeq
As this sequence of vectors $\{n^{(i)}, i \ge 0\}$ is nonnegative and monotonically decreasing, it converges
(in a finite number of iterations)
to the vector
\beq{codingA2266}
n^{(\infty)}=(\theta^c)^{*}(n).
\eeq
This leads to the following theorem.

\begin{figure}[ht]
	\centering
	\includegraphics[width=0.43\textwidth]{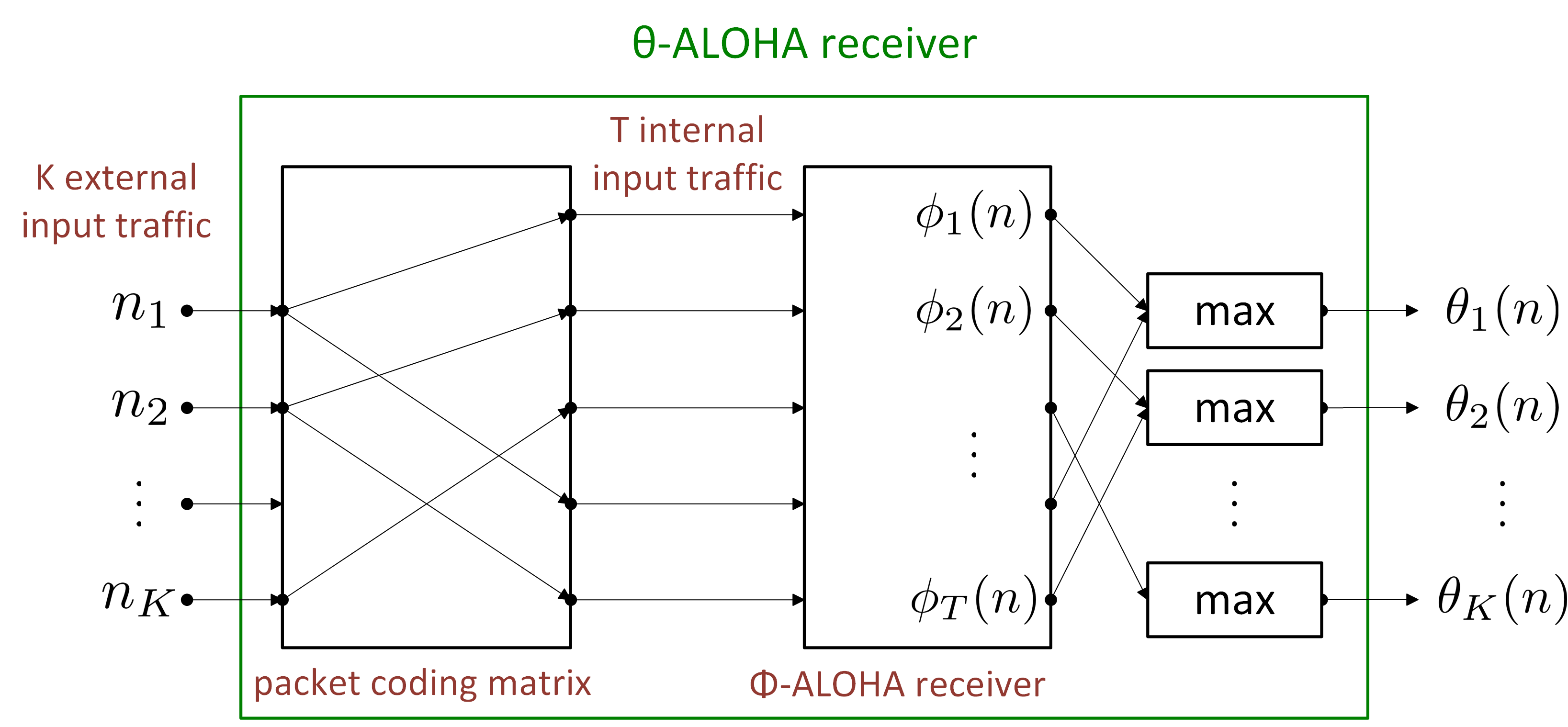}
	\caption{The $\theta$-ALOHA receiver in \req{codingA2222} for analyzing a system of ALOHA receivers with packet coding. In such a system, there are $K$ external classes of input traffic to
an all-or-nothing $\phi$-ALOHA receiver with $T$ internal classes of input traffic through a $K \times T$ packet coding matrix $H$.}
	\label{fig:codingA}
\end{figure}

\bthe{codingA} Consider a system with $K$ external classes of input traffic to
an all-or-nothing $\phi$-ALOHA receiver with $T$ internal classes of input traffic through a $K \times T$ packet coding matrix $H$ (see \rfig{codingA}). Then  the system is an all-or-nothing $((\theta^c)^*)^c$-ALOHA receiver,
where the function $\theta$ is defined in \req{codingA2222}.
\ethe

Interestingly, one can view the system of two cooperative ALOHA receivers in \rsec{coop}
as an ALOHA receiver with packet coding. First, an all-or-nothing $\phi$-ALOHA receiver with $K$ classes of input traffic and
an all-or-nothing $\psi$-ALOHA receiver with $K$ classes of input traffic can be viewed as an all-or-nothing $(\phi,\psi)$-ALOHA receiver with $2K$ classes of input traffic.
As the two cooperative ALOHA receivers are subject to the same $K$ classes of
external input,
the $K \times 2K$ coding matrix $H$ is simply a concatenation of two $K \times K$ identity matrices, i.e., both the first $K$ columns of $H$ and the last $K$ columns of $H$ are $K \times K$ identity matrices.
In view of \req{codingA2222}, we have
\beq{codingA2222b}
 \theta_k(n)=\max[\phi_t(n),\psi_t(n)],
 \eeq
and thus
\beq{codingA2222c}
\theta^c=\phi^c \wedge \psi^c.
\eeq
It then follows from \rlem{closure}(vii) that
\beq{codingA3333}
(\theta^c)^*= (\phi^c \wedge \psi^c)^* =(\phi^c \circ \psi^c)^*.
\eeq
The result in \rthe{codingA}  then recovers the result in \rthe{coop}.

We note that if the two receivers are not all-or-nothing ALOHA-receivers, then we have to identify the indices of the packets that are successfully received in order to carry out SIC decoding. As such, the identity in \req{codingA1111} is no longer valid as the class $k$ packets that are successfully received in $\phi_t(n^{(0)} H)$ for $t \in B_k$ might not be the same. One interesting example is to consider
the two cooperative ALOHA receivers in  \rsec{coop} with a single class of input traffic and
\beq{counter1111}
\phi(n)=\psi(n)=\left \{\begin{array}{cc}
		1 & \mbox{if}\; n =1\;\mbox{or}\;3\\
		0 & \mbox{otherwise}
	\end{array} \right ..
\eeq
When the system is subject to the deterministic load $n=3$, the decoding result from \rthe{coop} is
$$(\phi^c \circ \psi^c)^*(3)=2.$$
On the other hand, if we use the decoding method for packet coding and the packet that is successfully received by receiver 1 is different from that by receiver 2, then after the first SIC iteration, the number of packets remained to be decoded is reduced from 3 to 1.
After the second iteration, it will be further reduced to 0. This example shows that the decoding result by the method for packet coding
might be different from that from \rthe{coop} if the two receivers are not all-or-nothing ALOHA-receivers.

For  the general setting with joint
traffic multiplexing and packet coding, it can be represented by
a general $K\times T$ bipartite graph with a bi-adjacency matrix $H$. For such a bipartite graph, one can always add an intermediate stage of ${\tilde T}$ nodes  to form a tripartite graph such that (i) the first part is a $K \times {\tilde T}$ bipartite graph with a packet coding matrix $H_1$,
(ii) the second part is a ${\tilde T} \times T$ bipartite graph with a traffic multiplexing matrix $H_2$, and (iii) $H=H_1 H_2$. One simple way of doing this is to represent each edge in the general bipartite graph by a node in the intermediate stage.
By using the results for traffic multiplexing in \rthe{multiplex}  and packet coding in \rthe{codingA}, we can then analyze the setting with joint
traffic multiplexing and packet coding. We will further illustrate this by considering multiple cooperative $D$-fold ALOHA receivers in the next section.

\bsubsec{Multiple cooperative $D$-fold ALOHA receivers}{SAmul}

In this section, we show that  a system of multiple cooperative $D$-fold ALOHA receivers with joint
traffic multiplexing and packet coding is a $\phi$-ALOHA receiver. Moreover, the success function $\phi$ for such a system
can be computed by a max-sum message passing algorithm.



Consider a system
with $T$ cooperative $D$-fold ALOHA receivers and $K$ classes of input traffic through a $K \times T$  bi-adjacency matrix $H$.
The bi-adjacency matrix $H$ is a binary matrix that needs not to satisfy the constraints imposed on a traffic multiplexing matrix or a packet coding matrix.
Let $B_k$, $k=1,2, \ldots, K$, be the set of receivers associated with class $k$ input traffic (the set of nonzero elements in the $k^{th}$ row of $H$), and $C_t$, $t=1,2, \ldots, T$, be the set of
input traffic multiplexed into the $t^{th}$ receiver (the set of nonzero elements in the $t^{th}$ column of $H$).
To compute the success function for such a system subject to the deterministic load $n=(n_1, n_2, \ldots, n_K)$, we
 let $n^{(i)}=(n_1^{(i)}, n_2^{(i)}, \ldots, n_K^{(i)})$, where $n_k^{(i)}$ is the number of class $k$ packets remained to be decoded in the system after the $i^{th}$ SIC iteration, $k=1,2,\ldots, K$,  and $i=0,1,\ldots$. Clearly, $n^{(0)}=n$.
 Also, let $n_{k,t}^{(i)}$ be the number of class $k$ that are successfully received by the $t^{th}$ $D$-fold ALOHA receiver
 after the $i^{th}$ iteration.
For the first SIC iteration, we know from \req{multiplex1111} that $n_{k,t}^{(1)}=n_k^{(0)}$   if $k \in C_t$ and $\sum_{\ell \in C_t} n_\ell^{(0)} \le D$, and $n_{k,t}^{(1)}=0$ otherwise.
From \req{codingA1111}, we also know
 the number of class $k$ packets that are remained to be decoded after the first iteration is
\beq{codingA1111m}
n^{(1)}_k= n^{(0)}_k- \max_{t \in B_k}n_{k,t}^{(1)},
\eeq
 $k=1,2, \ldots, K$. In general, for the $i^{th}$ SIC iteration, we have
\beq{multiplex1111m}
n_{k,t}^{(i)}=\left \{\begin{array}{cc}
		n_k^{(i-1)} & \mbox{if}\; k \in C_t,\; \sum_{\ell \in C_t} n_\ell^{(i-1)} \le D \\
		0 & \mbox{otherwise}
	\end{array} \right .,
\eeq
 and
\beq{codingA1111m2}
n^{(i)}_k= n^{(i-1)}_k- \max_{t \in B_k}n_{k,t}^{(i)},
\eeq
 $k=1,2, \ldots, K$.
These two recursive equation leads to the max-sum message algorithm in Algorithm \ref{alg:maxsum}. Note that Steps 1 and 2 correspond to
the computation of $n_{k,t}^{(i)}$ in \req{multiplex1111m}, and Step 3 corresponds to the computation of $n^{(i)}_k$ in \req{codingA1111m2} (as the $D$-fold ALOHA receivers are all-or-nothing ALOHA receivers).

\begin{algorithm}\caption{The max-sum message passing algorithm}\label{alg:maxsum}

\noindent {\bf Input}  A $K\times T$ bi-adjacency $H$ and  a deterministic load $n=(n_1, n_2, \ldots, n_K)$.

\noindent {\bf Output} The number of class $k$ packets that are successfully received, $\phi_k(n)$, $k=1,2,\ldots, K$.

\noindent 0: Initially, set $n^{(0)}=n$ and $i=1$.


\noindent 1: For each class $k$ node, send a message $n_k^{(i-1)}$ to each receiver node in $B_k$.

\noindent 2: For each receiver $t$ node, compute the {\em sum} of the incoming messages.
 If the sum is not larger than $D$, return the original message to the sender. Otherwise,
return a message 0 to all the senders in $C_t$.

\noindent 3: For each class $k$ node, compute the {\em maximum} of the returning messages.
If the maximum is 0, set $n_k^{(i)}=n_k^{(i-1)}$. Otherwise, set $n_k^{(i)}=0$.

\noindent 4: If for every $k=1,2, \ldots, K$, the maximum of the returning messages is 0,
then no more packets can be successfully received. Set $\phi_k(n)=n_k -n_k^{(i)}$, $k=1,2, \ldots, K$.
Otherwise,
increase $i$ by 1 and repeat from Step 1.

\end{algorithm}

We note there is only a finite number of deterministic loads that need to be computed. To see this, note that if $n_k \ge D+1$ for some $k$, then there are at least $D+1$ packets transmitted to the $D$-fold ALOHA receivers in $B_k$. As such, no packets can be successfully received by the receivers in $B_k$, and we can remove all the edges connected to the receive nodes in $B_k$. Thus, the decoding results are the same for all
the deterministic loads with $n_k \ge D+1$ for some $k$. This implies that all the deterministic loads with $n_k > D+1$ is equivalent to $n_k=D+1$.
 As such, there are
$(D+2)^K$ equivalence classes (as we only need to consider $n_k=0,1,\ldots, D+1$ for each $k$).

Certainly, one can
use \req{phi1111} to compute the throughput of class $k$ users subject to a Poisson offered load $\rho=(\rho_1, \rho_2, \ldots, \rho_K)$ for such a $\phi$-ALOHA receiver. However,  it is more convenient to
compute the
throughput of class $k$ users from the probabilities of the $(D+2)^K$ equivalence classes.
Specifically, let
$p(n)$ is the probability of the deterministic load (equivalence class) $n$ subject to a Poisson offered load $\rho$.
For $d=0,1, \ldots, D$, let
\bearn
h_d(\rho_k)&=&\frac{e^{-\rho_k}\rho_k^d}{d!},\\
\eearn
and
$$h_{D+1}(\rho_k)=1- \sum_{d=0}^D \frac{e^{-\rho_k}\rho_k^d}{d!}.$$
For the Poisson distribution with mean $\rho_k$, the probabilities for $n_k$ are
$h_d(\rho_k)$, $d=0,1,\ldots, D+1$.
Since the Poisson offered loads from the $K$ classes are independent, we have
\beq{mulr2255}
p(n)=\prod_{k=1}^K h_{n_k}(\rho_k).
\eeq
Using \req{Poithrmul} yields
\beq{mulr3333}
 P_{{\rm suc},k}(\rhog)=\frac{S_k}{\rho_k}=\frac{1}{\rho_k}\sum_{n} \phi_k(n) \prod_{k=1}^K h_{n_k}(\rho_k).
 \eeq

\bex{phialoha2}{(Two cooperative $2$-fold ALOHA receivers)}
Consider the  system
with two cooperative $2$-fold ALOHA receivers and three classes of users.
Suppose that class 1 (resp. 2) packets are sent to receiver 1 (resp. 2), and class 3 packets are sent to
both receivers. The bi-adjacency matrix  of the association graph is
\beq{phialoha21}
H=
\left [ \begin{array}{ll}
1 & 0\\
0 & 1\\
1&  1
\end{array}
\right ].
\eeq
  As discussed in this section, it is a $\phi$-ALOHA receiver with the success function $\phi=(\phi_1,\phi_2,\phi_3)$ being specified in  Table \ref{table:phialoha}.
\eex

{
	\tiny
\begin{table}[htbp]
	\centering
\caption{The $\phi$-ALOHA receiver for a system with two cooperative $2$-fold ALOHA receivers in \rex{phialoha2}.}
			\begin{tabular}{|c|c|c|c|c|c|}
				\hline
$n_1$ & $n_2$ & $n_3$ & $\phi_1(n)$ & $\phi_2(n)$ & $\phi_3(n)$ \\ \hline
0 & 0 & 0 &  0 & 0 & 0 \\ \hline
0 & 0 & 1 &  0 & 0 & 1 \\ \hline
0 & 0 & 2 &  0 & 0 & 2 \\ \hline
0 & 0 & 3 &  0 & 0 & 0 \\ \hline
0 & 1 & 0 &  0 & 1 & 0 \\ \hline
0 & 1 & 1 &  0 & 1 & 1 \\ \hline
0 & 1 & 2 &  0 & 1 & 2 \\ \hline
0 & 1 & 3 &  0 & 0 & 0 \\ \hline
0 & 2 & 0 &  0 & 2 & 0 \\ \hline
0 & 2 & 1 &  0 & 2 & 1 \\ \hline
0 & 2 & 2 &  0 & 2 & 2 \\ \hline
0 & 2 & 3 &  0 & 0 & 0 \\ \hline
0 & 3 & 0 &  0 & 0 & 0 \\ \hline
0 & 3 & 1 &  0 & 0 & 1 \\ \hline
0 & 3 & 2 &  0 & 0 & 2 \\ \hline
0 & 3 & 3 &  0 & 0 & 0 \\ \hline
1 & 0 & 0 &  1 & 0 & 0 \\ \hline
1 & 0 & 1 &  1 & 0 & 1 \\ \hline
1 & 0 & 2 &  1 & 0 & 2 \\ \hline
1 & 0 & 3 &  0 & 0 & 0 \\ \hline
1 & 1 & 0 &  1 & 1 & 0 \\ \hline
1 & 1 & 1 &  1 & 1 & 1 \\ \hline
1 & 1 & 2 &  0 & 0 & 0 \\ \hline
1 & 1 & 3 &  0 & 0 & 0 \\ \hline
1 & 2 & 0 &  1 & 2 & 0 \\ \hline
1 & 2 & 1 &  1 & 2 & 1 \\ \hline
1 & 2 & 2 &  0 & 0 & 0 \\ \hline
1 & 2 & 3 &  0 & 0 & 0 \\ \hline
1 & 3 & 0 &  1 & 0 & 0 \\ \hline
1 & 3 & 1 &  1 & 0 & 1 \\ \hline
1 & 3 & 2 &  0 & 0 & 0 \\ \hline
1 & 3 & 3 &  0 & 0 & 0 \\ \hline

2 & 0 & 0 &  2 & 0 & 0 \\ \hline
2 & 0 & 1 &  2 & 0 & 1 \\ \hline
2 & 0 & 2 &  2 & 0 & 2 \\ \hline
2 & 0 & 3 &  0 & 0 & 0 \\ \hline
2 & 1 & 0 &  2 & 1 & 0 \\ \hline
2 & 1 & 1 &  2 & 1 & 1 \\ \hline
2 & 1 & 2 &  0 & 0 & 0 \\ \hline
2 & 1 & 3 &  0 & 0 & 0 \\ \hline
2 & 2 & 0 &  2 & 2 & 0 \\ \hline
2 & 2 & 1 &  0 & 0 & 0 \\ \hline
2 & 2 & 2 &  0 & 0 & 0 \\ \hline
2 & 2 & 3 &  0 & 0 & 0 \\ \hline
2 & 3 & 0 &  2 & 0 & 0 \\ \hline
2 & 3 & 1 &  0 & 0 & 0 \\ \hline
2 & 3 & 2 &  0 & 0 & 0 \\ \hline
2 & 3 & 3 &  0 & 0 & 0 \\ \hline
3 & 0 & 0 &  0 & 0 & 0 \\ \hline
3 & 0 & 1 &  0 & 0 & 1 \\ \hline
3 & 0 & 2 &  0 & 0 & 2 \\ \hline
3 & 0 & 3 &  0 & 0 & 0 \\ \hline
3 & 1 & 0 &  0 & 1 & 0 \\ \hline
3 & 1 & 1 &  0 & 1 & 1 \\ \hline
3 & 1 & 2 &  0 & 0 & 0 \\ \hline
3 & 1 & 3 &  0 & 0 & 0 \\ \hline
3 & 2 & 0 &  0 & 2 & 0 \\ \hline
3 & 2 & 1 &  0 & 0 & 0 \\ \hline
3 & 2 & 2 &  0 & 0 & 0 \\ \hline
3 & 2 & 3 &  0 & 0 & 0 \\ \hline
3 & 3 & 0 &  0 & 0 & 0 \\ \hline
3 & 3 & 1 &  0 & 0 & 0 \\ \hline
3 & 3 & 2 &  0 & 0 & 0 \\ \hline
3 & 3 & 3 &  0 & 0 & 0 \\ \hline
			\end{tabular}
	\label{table:phialoha}
\end{table}
}

\bsec{Numerical results}{num}

\bsubsec{Two cooperative  $D$-fold ALOHA  receivers with  URLLC traffic and eMBB traffic}{mulrec}

\begin{figure}[ht]
	\centering
	\includegraphics[width=0.40\textwidth]{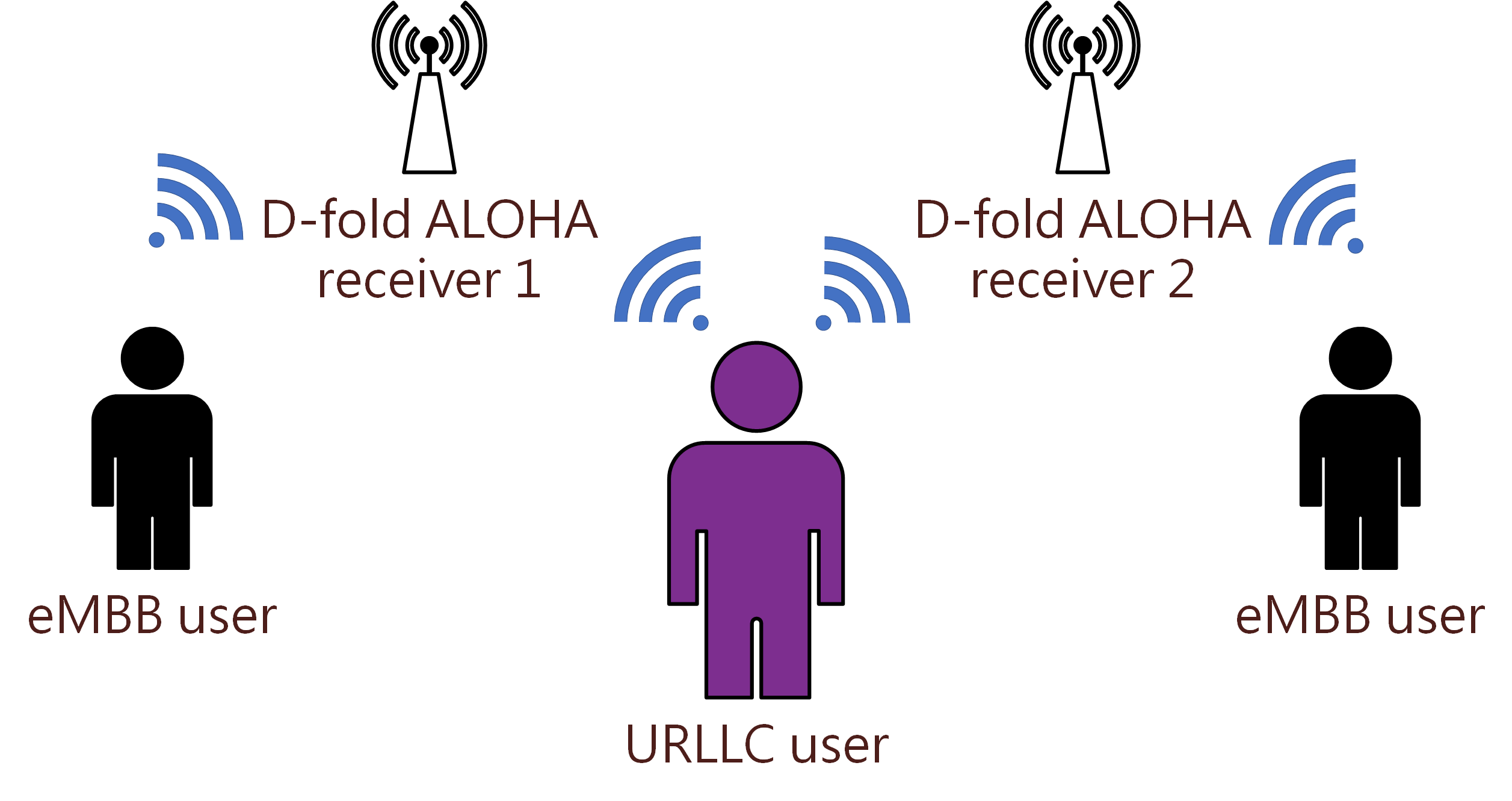}
	\caption{An illustration for two cooperative $D$-fold ALOHA receivers with two classes of external input traffic: URLLC packets are multicast to both receivers, while eMBB packets can only be routed to one of the two receivers.}
	\label{fig:difrec2}
\end{figure}

In this section, we demonstrate how the framework of Poisson receivers and the framework of ALOHA receivers can be used for providing differentiated services between URLLC traffic and eMBB traffic in uplink transmissions.

\subsubsection{Wireless channel model}
\label{sec:wireless}

We consider a wireless channel that is modelled by two cooperative $D$-fold ALOHA receivers  in \rex{phialoha2}.
There are three classes of input traffic.
Class 1 (resp. 2) packets are sent to the first (resp. 2) receiver, and
class 3  packets are sent to both receivers (see \rfig{difrec2} for an illustration).
For such a channel model, it is a $\phi$-ALOHA receiver with the success function $\phi$ specified in  Table \ref{table:phialoha}.
By using \req{mulr3333}, we can compute the success probability functions $P_{{\rm suc},k}(\rho_1,\rho_2,\rho_3)$, $k=1,2,3$, of the induced Poisson receiver.
The first two classes are
 eMBB traffic and the third class is URLLC traffic. Each eMBB packet can be routed to  a class 1 packet or a class 2 packet with an equal probability. As a result of the inverse multiplexer in Example 3 of \cite{chang2020Poisson}, one can use \rthe{routing} to model such a channel as a Poisson receiver with two classes of external input traffic, URLLC traffic (external class 1) and eMBB traffic (external class 2).
The success probability functions of these two external classes are
\bear{invm4444r}
&&\tilde P_{{\rm suc},1}(\tilde \rho_1,\tilde \rho_2)=P_{{\rm suc},3}(\tilde \rho_2/2,\tilde \rho_2/2,\tilde \rho_1) \nonumber \\
&&\tilde P_{{\rm suc},2}(\tilde \rho_1,\tilde \rho_2)=\frac{1}{2} P_{{\rm suc},1}(\tilde \rho_2/2,\tilde \rho_2/2,\tilde \rho_1 )\nonumber\\
&&\quad\quad\quad+ \frac{1}{2}P_{{\rm suc},2}(\tilde \rho_2/2,\tilde \rho_2/2,\tilde \rho_1).
\eear

\subsubsection{Coded random access for a use case}
\label{sec:usecase}

To provide differentiated services between URLLC traffic and eMBB traffic over the wireless channel model in Section \ref{sec:wireless}, we consider the particular use case in Section IV.C of \cite{chang2020Poisson} for supporting precise cooperative robotic motion control defined in use case 1 of mobile robots in \cite{3gpp.22.104}. For this use case, the message (packet) size of URLLC traffic is 40 bytes, and the transmission time interval (TTI) is 1 ms.
According to Table 4.1A-2 in \cite{3gpp.36.306}, the maximum number of bits of an uplink shared channel (UL-SCH) transport block transmitted within a TTI is 105,528 bits (i.e., 13,191 bytes). As a conservative design, the number of packet transmissions (minislots) within one TTI  is set to be 256. As there can be as many as $D$ successful packet transmissions in a $D$-fold ALOHA receiver, we set
$T=256/D$. Thus, for each TTI, there are $T$ independent Poisson receivers for the wireless channel model in Section \ref{sec:wireless}. The number of active URLLC users $N_1=G_1T$ is set to be 50, and each URLLC user uses a coded random access scheme that transmits its packet for $L=5$ times uniformly and independently to the $T$ Poisson receivers.
On the other hand, each eMBB user is scheduled to transmit exactly one packet in a (randomly assigned) time slot in each TTI.
The packets transmitted by URLLC users are superposed with the scheduled eMBB traffic \cite{anand2020joint} over the wireless channel in Section \ref{sec:wireless}.
By viewing each minislot as a Poisson receiver with two classes of input traffic in Section \ref{sec:wireless},
the coded random access scheme corresponds to a system of Poisson receivers with packet coding in \rthe{coding}, where
the degree distribution of URLLC traffic is $\Lambda_1(x)=x^L$ with $L =5$ (see \req{mean0000mul}), and the degree distribution of eMBB traffic is $\Lambda_2(x)=x$. We note that  URLLC transmissions in such a system are decoded first by using the SIC technique. eMBB transmissions that overlap with undecodable URLLC transmissions can be treated as {\em erased} or {\em punctured} \cite{popovski20185g} and can be protected by using another layer of error correction codes.

We are interested in addressing the question of how many eMBB users $N_2=G_2T$ can be admitted to the system so that the packet error probability of URLLC traffic is less than $10^{-5}$.



\begin{figure}[ht]
	\centering
	\includegraphics[width=0.40\textwidth]{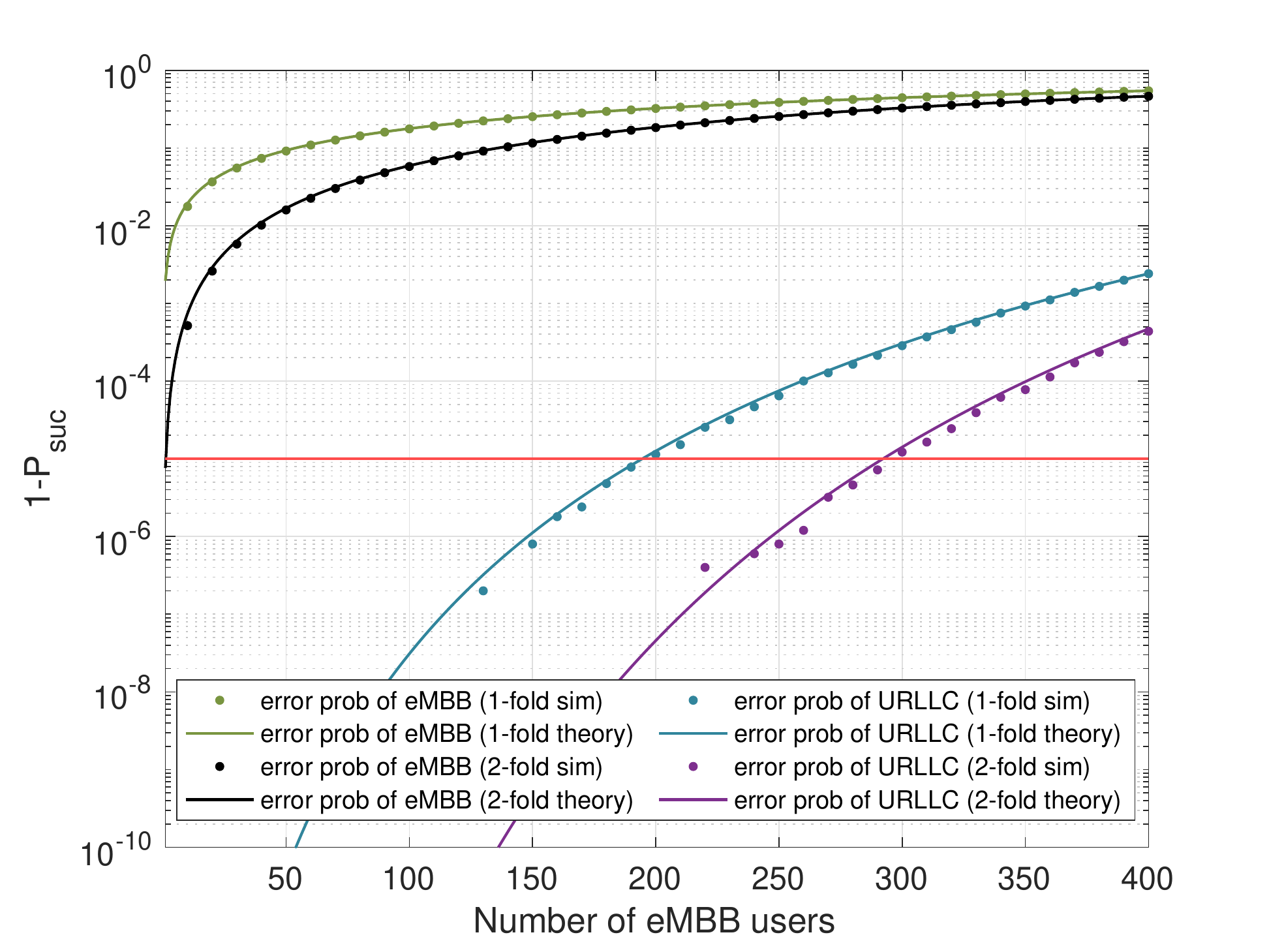}
	\caption{The effect of the number of eMBB users on the error probability of URLLC users in a system of two cooperative $D$-fold ALOHA receivers for $D=1$ and $D=2$.}
	\label{fig:difrec_result}
\end{figure}

In \rfig{difrec_result}, we show the theoretical results and the simulation results for $D=1$ and $D=2$.
Each data point for the estimated error probability is obtained by averaging over 100,000 independent runs.
Also, we set the number of iterations for SIC to be 100 ($i=100$).  As shown in \rfig{difrec_result}, our theoretical results match extremely well with the simulation results except for those data points with extremely small error probabilities. For $D=1$, the 1-fold ALOHA receiver is the simply the SA system considered in Section IV.C of \cite{chang2020Poisson}. One can see from this figure that the performance of the 2-fold ALOHA receivers is significantly better than that of the SA system. In particular, the number of eMBB users can be admitted to the system is increased from 194 for the system with two 1-fold ALOHA receivers to 292 for the system with two 2-fold ALOHA receivers (while keeping the error probability of URLLC packets lower than $10^{-5}$).


\bsubsec{Rayleigh  block fading channel with  URLLC traffic and eMBB traffic}{rayleigh}

\begin{figure}[ht]
	\centering
	\includegraphics[width=0.40\textwidth]{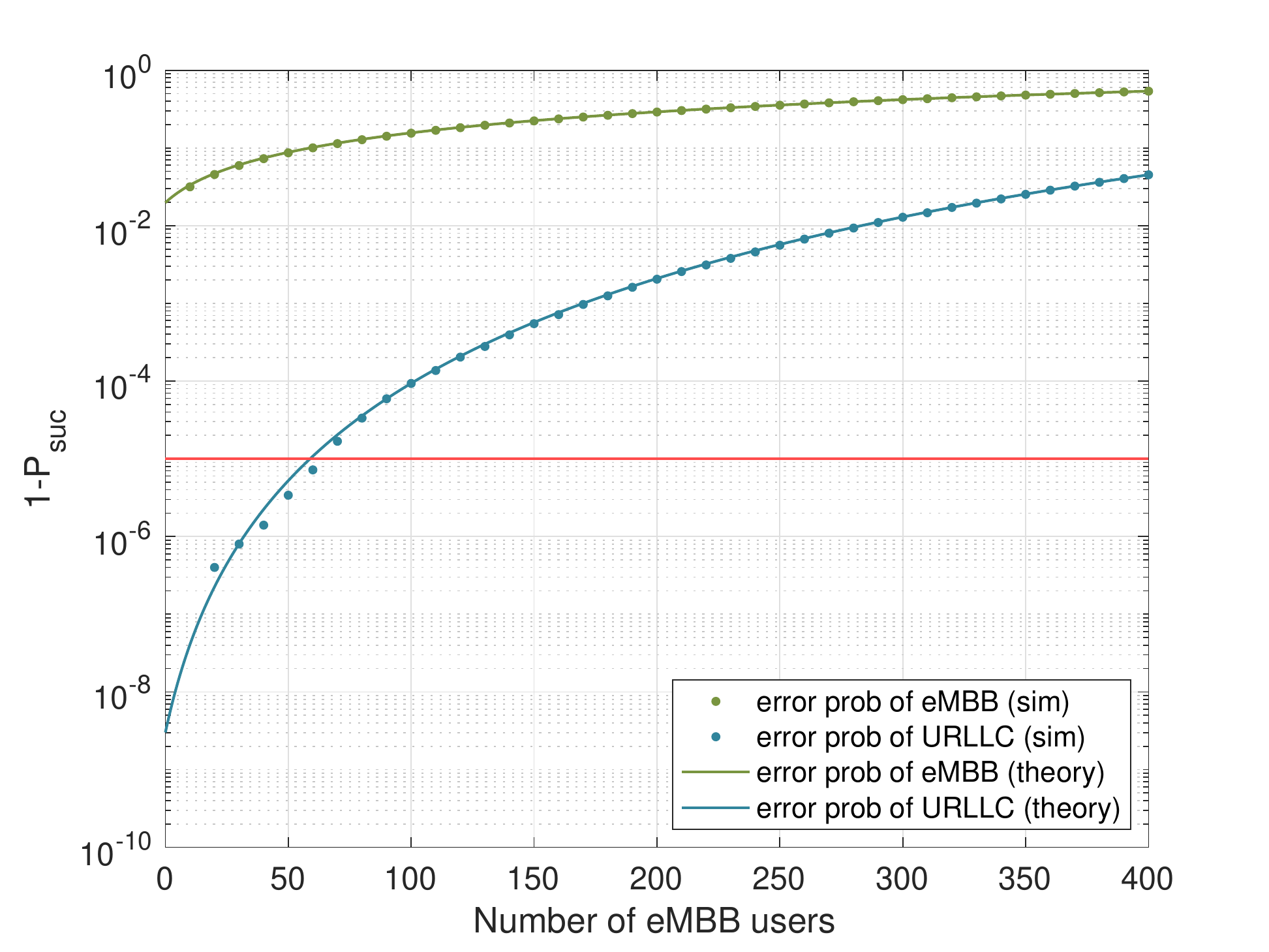}
	\caption{The effect of the number of eMBB users on the error probability of URLLC users in the Rayleigh  block fading channel with capture.}
	\label{fig:rayleigh}
\end{figure}

In this section, we consider the Rayleigh  block fading channel with capture in \rsec{fading}.
Such a wireless channel can be modelled by a Poisson receiver with a single class of input in \req{fading8888}.
By using \rthe{routing} for Poisson receivers with packet routing, such a wireless channel can be extended  to a
Poisson receiver with two classes of input traffic.
The success probability functions for these two classes of input traffic are as follows:
\bear{fading88882}
&&P_{{\rm suc},1}(\rho_1,\rho_2)=P_{{\rm suc},2}(\rho_1,\rho_2)\nonumber \\
&&=\sum_{t=0}^\infty  \sum_{\rone=0}^{t} \frac{e^{-(\rho_1+\rho_2)} (\rho_1+\rho_2)^t}{(t-\rone)!} \frac{e^{-\frac{1}{\gamma}((1+b)^{\rone+1}-1)}}{(1+b)^{(\rone+1)\left(t-\frac{\rone}{2}\right)}}.\nonumber \\
\eear
To provide differentiated services between URLLC traffic and eMBB traffic, we use the coded random access scheme with $T$ minislots  as described in Section \ref{sec:usecase}. We assume that the channel gains for these two  classes in $T$ minislots are {\em independent}, and thus such a system corresponds to a system of Poisson receivers with packet coding in \rthe{coding}. In our experiments, we use the same parameters as those in Section \ref{sec:usecase}, i.e., the number of minislots $T=256$, the number of URLLC users $N_1=50$, the number of SIC iterations $i=100$,
 the degree distribution of URLLC traffic $\Lambda_1(x)=x^5$, and the degree distribution of eMBB traffic $\Lambda_2(x)=x$. For the Rayleigh  block fading channel, we set $\gamma=20dB$ and $b=3dB$.
As in Section \ref{sec:usecase}, each data point is obtained by averaging over 100,000 independent runs. In \rfig{rayleigh}, we show the theoretical results and the simulation results for the effect of the number of eMBB users on the error probability of URLLC users in the Rayleigh  block fading channel with capture.
Once again, our theoretical results match extremely well with the simulation results.
The number of eMBB users can be admitted to the system is roughly 58 while keeping the error probability of URLLC packets lower than $10^{-5}$.

\bsec{Conclusion}{con}

In this paper, we developed a deterministic framework of $\phi$-ALOHA receivers that can be incorporated into the probabilistic framework of Poisson receivers for
analyzing coded multiple access with SIC. Like the theory of network calculus, there are various algebraic properties for several operations on functions, including minimum $\wedge$, composition $\circ$, closure $*$, and complement $c$.
As such, small ALOHA receivers can be used as building blocks for constructing a large ALOHA receiver.
In particular, we showed various closure properties of ALOHA receivers, including (i) ALOHA receivers in tandem,
(ii) cooperative receivers, (iii) ALOHA receivers with traffic multiplexing, and (iv) ALOHA receivers with packet coding. As an illustrating example, we computed/simulated the numerical results of a system that uses two cooperative $D$-fold ALOHA receivers with packet coding to provide differentiated services between URLLC traffic and eMBB traffic.
The theoretical results in this example match extremely well with the simulation results (except for those data points with extremely small error probabilities).

\begin{IEEEbiography}[{\includegraphics[width=1in,height=1.25in,clip,keepaspectratio]{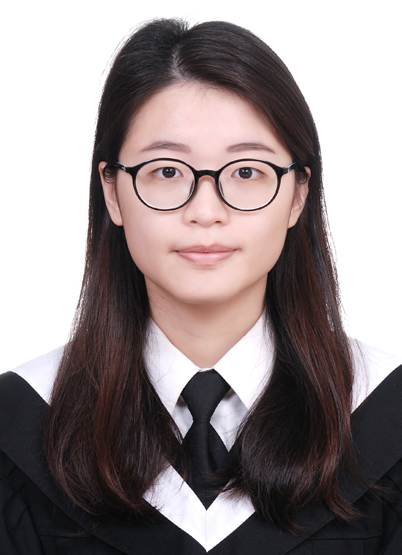}}]
	{Tzu-Hsuan Liu} received the B.S. degree in communication engineering from National Central University, Taoyuan, Taiwan (R.O.C.), in 2018. She is currently pursuing the M.S. degree in the Institute of Communications Engineering, National Tsing Hua University, Hsinchu, Taiwan (R.O.C.). Her research interest is in 5G wireless communication.
\end{IEEEbiography}

\begin{IEEEbiography}[{\includegraphics[width=1in,height=1.25in,clip,keepaspectratio]{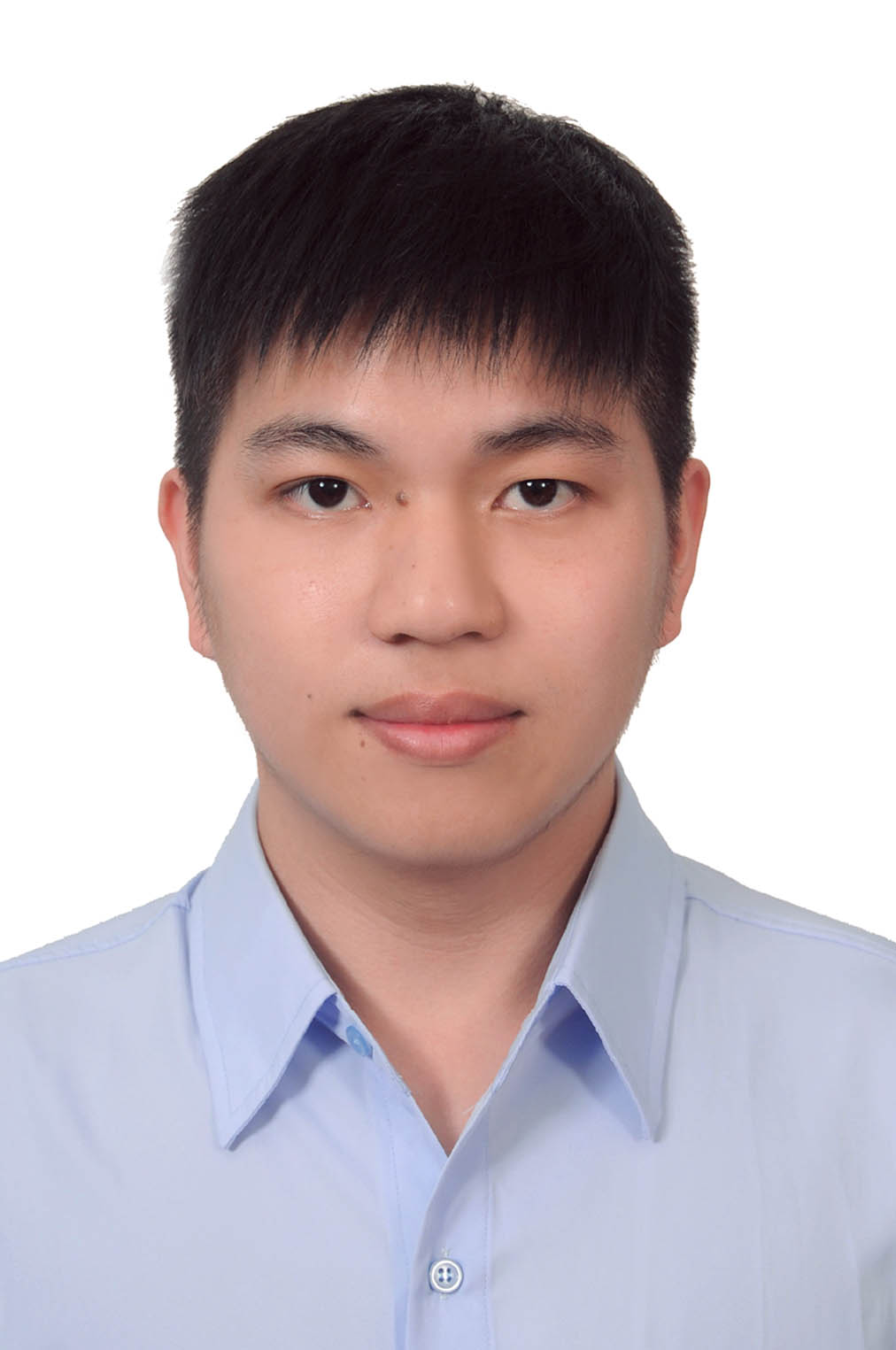}}]
{Che-Hao Yu} received his B.S. degree in mathematics from National Tsing-Hua University, Hsinchu, Taiwan (R.O.C.), in 2018, and the M.S. degree in communications engineering from National Tsing Hua University, Hsinchu, Taiwan (R.O.C.), in 2020. His research interest is in 5G wireless communication.
\end{IEEEbiography}

\begin{IEEEbiography}[{\includegraphics[width=1in,height=1.25in,clip,keepaspectratio]{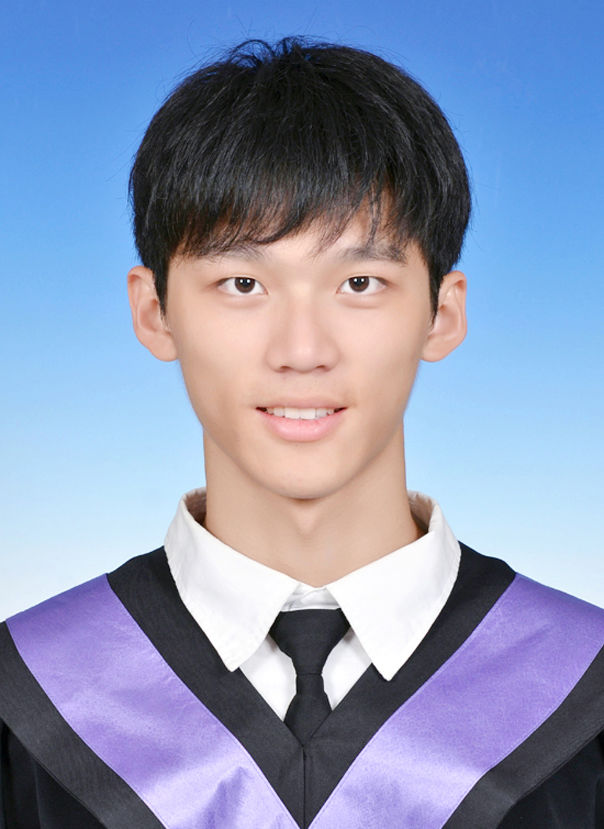}}]
{Yi-Jheng Lin} received his B.S. degree in electrical engineering from National Tsing Hua University, Hsinchu, Taiwan, in 2018. He is currently pursuing the Ph.D. degree in the Institute of Communications Engineering, National Tsing Hua University, Hsinchu, Taiwan. His research interests include wireless communication and cognitive radio networks.
\end{IEEEbiography}

\begin{IEEEbiography}[{\includegraphics[width=1in,height=1.25in,clip,keepaspectratio]{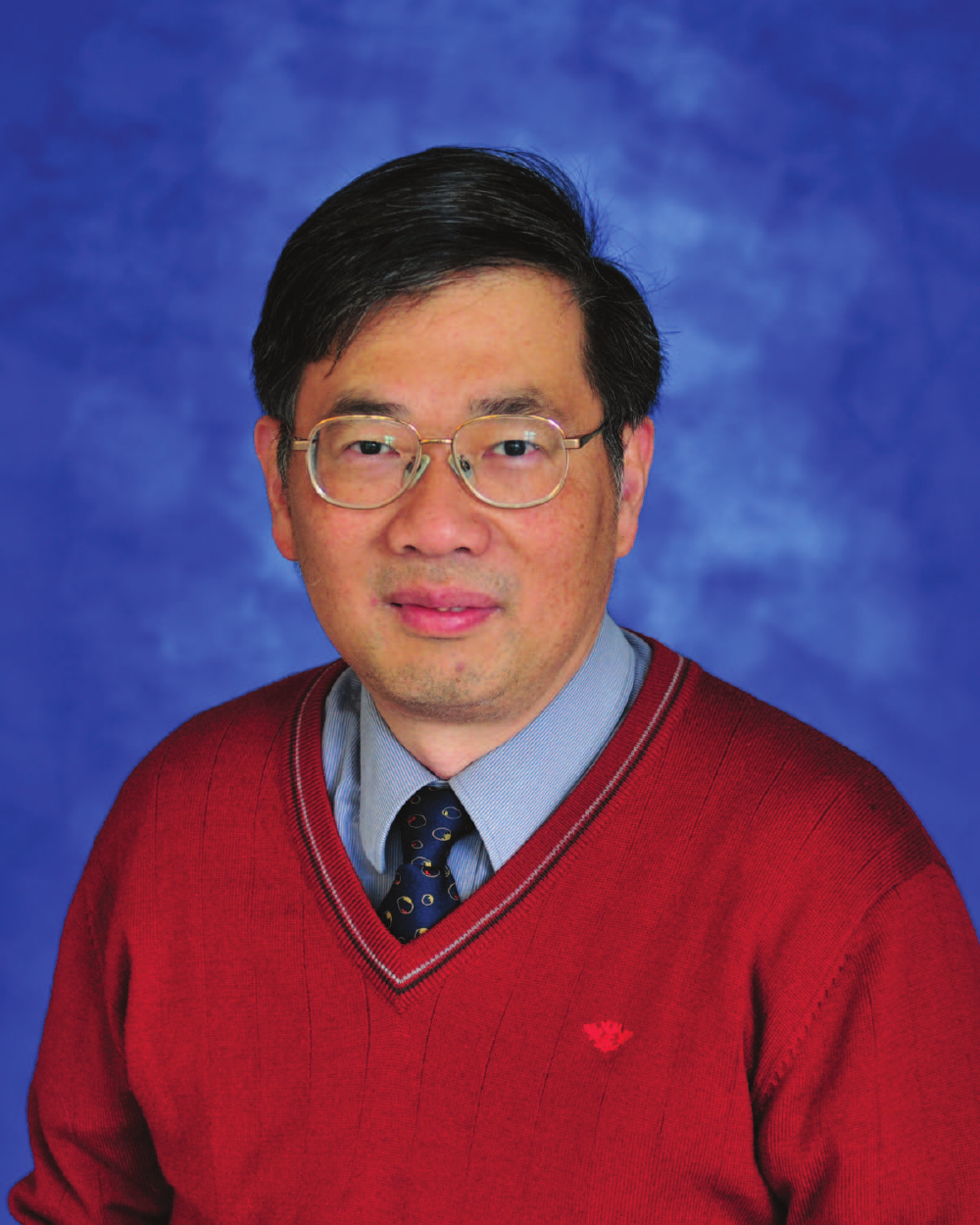}}]
	{Cheng-Shang Chang}
	(S'85-M'86-M'89-SM'93-F'04)
	received the B.S. degree from National Taiwan
	University, Taipei, Taiwan, in 1983, and the M.S.
	and Ph.D. degrees from Columbia University, New
	York, NY, USA, in 1986 and 1989, respectively, all
	in electrical engineering.
	
	From 1989 to 1993, he was employed as a
	Research Staff Member with the IBM Thomas J.
	Watson Research Center, Yorktown Heights, NY,
	USA. Since 1993, he has been with the Department
	of Electrical Engineering, National Tsing Hua
	University, Taiwan, where he is a Tsing Hua Distinguished Chair Professor. He is the author
	of the book Performance Guarantees in Communication Networks (Springer,
	2000) and the coauthor of the book Principles, Architectures and Mathematical
	Theory of High Performance Packet Switches (Ministry of Education, R.O.C.,
	2006). His current research interests are concerned with network science, big data analytics,
	mathematical modeling of the Internet, and high-speed switching.
	
	Dr. Chang served as an Editor for Operations Research from 1992 to 1999,
	an Editor for the {\em IEEE/ACM TRANSACTIONS ON NETWORKING} from 2007
	to 2009, and an Editor for the {\em IEEE TRANSACTIONS
		ON NETWORK SCIENCE AND ENGINEERING} from 2014 to 2017. He is currently serving as an Editor-at-Large for the {\em IEEE/ACM
		TRANSACTIONS ON NETWORKING}. He is a member of IFIP Working
	Group 7.3. He received an IBM Outstanding Innovation Award in 1992, an
	IBM Faculty Partnership Award in 2001, and Outstanding Research Awards
	from the National Science Council, Taiwan, in 1998, 2000, and 2002, respectively.
	He also received Outstanding Teaching Awards from both the College
	of EECS and the university itself in 2003. He was appointed as the first Y. Z.
	Hsu Scientific Chair Professor in 2002. He received the Merit NSC Research Fellow Award from the
	National Science Council, R.O.C. in 2011. He also received the Academic Award in 2011 and the National Chair Professorship in 2017 from
	the Ministry of Education, R.O.C. He is the recipient of the 2017 IEEE INFOCOM Achievement Award.
\end{IEEEbiography}

\begin{IEEEbiography}
	[{\includegraphics[width=1in,height=1.25in,clip,keepaspectratio]{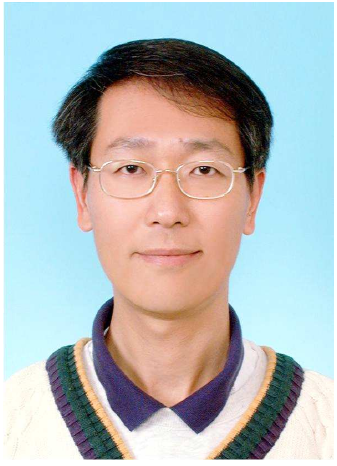}}]
	{Duan-Shin Lee}(S'89-M'90-SM'98) received the B.S. degree from National Tsing Hua
	University, Taiwan, in 1983, and the MS and Ph.D. degrees from
	Columbia University, New York, in 1987 and 1990, all in electrical
	engineering.  He worked as a research staff member at the C\&C Research Laboratory
	of NEC USA, Inc. in Princeton, New Jersey from 1990 to 1998.  He joined the
	Department of Computer Science of National Tsing Hua University in Hsinchu,
	Taiwan, in 1998.  Since August 2003, he has been a professor.  He received
	a best paper award from the Y.Z. Hsu Foundation in 2006.  He served as
	an editor for the Journal of Information Science and Engineering between
	2013 and 2015.  He is currently an editor for Performance Evaluation.
	Dr. Lee's current research interests are network science, game theory,
	machine learning and high-speed networks.  He is a senior IEEE member.
\end{IEEEbiography}

\clearpage

\section*{Appendix A}

\setcounter{section}{1}

It this section, we show the derivation of \req{fading5555}.

\begin{align*}
&\mathrm{Pr}\left\{\dfrac{X_1}{\sum_{j=2}^{N}X_j+\frac{1}{\gamma}}\geq b, \dfrac{X_2}{\sum_{j=3}^{N}X_j+\frac{1}{\gamma}}\geq b, \cdots \dfrac{X_{r}}{\sum_{j=r+1}^{N}X_j+\frac{1}{\gamma}}\geq b\right\} \\
= & \int_0^\infty dx_{N}\cdots \int_0^\infty dx_{r+1}\times \int_{b(\sum_{j=r+1}^{N}x_j+\frac{1}{\gamma})}^\infty dx_r\cdots \int_{b(\sum_{j=2}^{N}x_j+\frac{1}{\gamma})}^\infty dx_1\times e^{-x_{N}}\cdots e^{-x_1}  \\
= & \int_0^\infty dx_{N}\cdots \int_0^\infty dx_{r+1}\times \int_{b(\sum_{j=r+1}^{N}x_j+\frac{1}{\gamma})}^\infty dx_r\cdots \int_{b(\sum_{j=2}^{N}x_j+\frac{1}{\gamma})}^\infty e^{-x_1}dx_1\times e^{-\sum_{j=2}^{N} x_j} \\
= & \int_0^\infty dx_{N}\cdots \int_0^\infty dx_{r+1}\times \int_{b(\sum_{j=r+1}^{N}x_j+\frac{1}{\gamma})}^\infty dx_r\cdots \int_{b(\sum_{j=3}^{N}x_j+\frac{1}{\gamma})}^\infty dx_2\times e^{-b\sum_{j=2}^{N}x_j-b\frac{1}{\gamma}}e^{-\sum_{j=2}^{N} x_j}  \\
= & e^{-b\frac{1}{\gamma}}\times \int_0^\infty dx_{N}\cdots \int_0^\infty dx_{r+1}\times \int_{b(\sum_{j=r+1}^{N}x_j+\frac{1}{\gamma})}^\infty dx_r\cdots \int_{b(\sum_{j=3}^{N}x_j+\frac{1}{\gamma})}^\infty e^{-(1+b)x_2} dx_2\times e^{-(1+b)\sum_{j=3}^{N}x_j}  \\
= & e^{-b\frac{1}{\gamma}}\frac{e^{-b(1+b)\frac{1}{\gamma}}}{(1+b)}\times \int_0^\infty dx_{N}\cdots \int_0^\infty dx_{r+1}\times \\
& \int_{b(\sum_{j=r+1}^{N}x_j+\frac{1}{\gamma})}^\infty dx_r\cdots \int_{b(\sum_{j=4}^{N}x_j+\frac{1}{\gamma})}^\infty dx_3\times e^{-(1+b)\sum_{j=3}^{N}x_j}e^{-(1+b)b\sum_{j=3}^{N}x_j} \\
= & e^{-b\frac{1}{\gamma}}\frac{e^{-b(1+b)\frac{1}{\gamma}}}{(1+b)}\times \int_0^\infty dx_{N}\cdots \int_0^\infty dx_{r+1}\times \\
& \int_{b(\sum_{j=r+1}^{N}x_j+\frac{1}{\gamma})}^\infty dx_{r}\cdots \int_{b(\sum_{j=4}^{N}x_j+\frac{1}{\gamma})}^\infty e^{-(1+b)^2x_3}dx_3\times e^{-(1+b)^2\sum_{j=4}^{N}x_j} \\
=& e^{-b\frac{1}{\gamma}}\frac{e^{-b(1+b)\frac{1}{\gamma}}}{(1+b)}\frac{e^{-b(1+b)^2\frac{1}{\gamma}}}{(1+b)^2}\times \int_0^\infty dx_{N}\cdots \int_0^\infty dx_{r+1}\times \\
& \int_{b(\sum_{j=r+1}^{N}x_j+\frac{1}{\gamma})}^\infty dx_{r}\cdots \int_{b(\sum_{j=5}^{N}x_j+\frac{1}{\gamma})}^\infty e^{-(1+b)^3x_4}dx_4\times e^{-(1+b)^3\sum_{j=5}^{N}x_j} \\
\vdots & \\
= &
\prod_{k=0}^{r-2}\left(\frac{e^{-b\frac{1}{\gamma}(1+b)^k}}{(1+b)^k}\right) \times \int_0^\infty dx_{N}\cdots \int_0^\infty dx_{r+1}\times \int_{b(\sum_{j=r+1}^{N}x_j+\frac{1}{\gamma})}^\infty e^{-(1+b)^{r}x_{r}} dx_{r}\times e^{-(1+b)^r\sum_{j=r+1}^{N}x_j} \\
= &
\prod_{k=0}^{r-1}\left(\frac{e^{-b\frac{1}{\gamma}(1+b)^k}}{(1+b)^k}\right) \times \int_0^\infty dx_{N}\cdots \int_0^\infty dx_{r+1}\times e^{-(1+b)^{r}\sum_{j=r+1}^{N}x_j} \\
= &
\prod_{k=0}^{r-1}\left(\frac{e^{-b\frac{1}{\gamma}(1+b)^k}}{(1+b)^k}\right) \times \prod_{k=r+1}^{N}\left(\int_0^\infty e^{-(1+b)^{r}x_k}dx_k\right)
=
\prod_{k=0}^{r-1}\left(\frac{e^{-b\frac{1}{\gamma}(1+b)^k}}{(1+b)^k}\right) \times \left(\frac{1}{(1+b)^{r}}\right)^{N-r} \\
= &
\left(\frac{e^{-b\frac{1}{\gamma}\sum_{k=0}^{r-1}(1+b)^k}}{(1+b)^{\sum_{k=0}^{r-1}k}}\right) \times \left(\frac{1}{(1+b)^{r}}\right)^{N-r}
=
\left(\frac{e^{-\frac{1}{\gamma}((1+b)^{r}-1)}}{(1+b)^{\frac{r(r-1)}{2}}}\right) \times \left(\frac{1}{(1+b)^{r}}\right)^{N-r} \\
= & \frac{e^{-\frac{1}{\gamma}((1+b)^{r}-1)}}{(1+b)^{\frac{r(r-1)}{2}+r(N-r)}}
=  \frac{e^{-\frac{1}{\gamma}((1+b)^{r}-1)}}{(1+b)^{r\left(N-1-\frac{r-1}{2}\right)}}\\
\end{align*}

\end{document}